\documentclass[aps,pra,amsmath,amssymb,amsfonts,superscriptaddress,lengthcheck,twocolumn,longbibliography]{revtex4-1}

\usepackage{dcolumn}    
\usepackage{graphicx}
\usepackage{amsmath, bbm}
\usepackage{latexsym}
\usepackage{amsfonts}   
\usepackage{amssymb}
\usepackage{array}      
\usepackage{epsfig}
\usepackage{txfonts}
\usepackage{xcolor,braket}
\usepackage[colorlinks=true,linkcolor=blue,urlcolor=blue,citecolor=blue,pdfusetitle]{hyperref}
\usepackage{hyperref}
\usepackage{ulem}
\usepackage{soul}
\usepackage{dsfont}
\usepackage{physics}
\usepackage{bm}
\usepackage{tensor}

\usepackage{relsize}

\usepackage{mathrsfs}

\usepackage{cancel}
\usepackage{enumitem}

\newcommand{\F}{\mathcal{F}}

\begin{document}
\title{Bounding fidelity in quantum feedback control: theory and applications to Dicke state preparation} 
       
\author{Eoin O'Connor}
\affiliation{Dipartimento di Fisica “Aldo Pontremoli”, Università degli Studi di Milano, I-20133 Milan, Italy}
\author{Hailan Ma}
\affiliation{School of Engineering, Australian National University, Canberra, ACT 2601, Australia}
\author{Marco G. Genoni}
\affiliation{Dipartimento di Fisica “Aldo Pontremoli”, Università degli Studi di Milano, I-20133 Milan, Italy}
\begin{abstract}
Achieving unit fidelity in quantum state preparation is often impossible in the presence of environmental decoherence. While continuous monitoring and feedback control can improve fidelity, perfect state preparation remains elusive in many scenarios. Inspired by quantum speed limits, we derive a fundamental bound on the steady-state average fidelity achievable via continuous monitoring and feedback control. This bound depends only on the unconditional Lindblad dynamics, the Hamiltonian variance, and the target state. We also adapt the bound to the case of Markovian feedback strategies. We then focus on preparing Dicke states in an atomic ensemble subject to collective damping and dispersive coupling. By imposing additional constraints on control Hamiltonians and monitoring strategies, we derive tighter fidelity bounds. Finally, we propose specific control strategies and validate them using reinforcement learning. Benchmarking their performance against our theoretical bounds highlights the relevance and usefulness of these bounds in characterizing quantum feedback control strategies.
\end{abstract}
\date{\today}

\maketitle

\section{Introduction}
Quantum state engineering plays a fundamental role in both theoretical and applied quantum science. From a theoretical standpoint, it provides a platform to investigate key questions in quantum physics, particularly the preparation of states that exhibit non-classical behavior. On the practical side, it serves as a cornerstone for developing future quantum technologies, including quantum communication, computation, and metrology. 
Quantum control theory offers techniques for deterministic state preparation in idealized closed quantum systems~\cite{Koch2022_QOC}. However, in real experimental scenarios, environmental noise is inevitable. Unless a target state is inherently robust to such noise, achieving deterministic preparation of the desired state becomes challenging. In this work, we focus on the framework of quantum feedback control~\cite{wisemanQuantumMeasurementControl2009}, where continuous monitoring of the environment enables real-time adjustments to the system based on acquired information. This approach has been shown to be effective not only for driving a system toward a specific quantum state~\cite{Doherty1999,wang2001feedback,stockton2004deterministic,geremia2006deterministic,yanagisawa2006quantum,mirrahimi2007stabilizing,yamamoto2007feedback,negretti2007quantum,bouten2009discrete,Patti2017,fosel2018reinforcement,Saiphet2021,Borah2021,porotti2022deep,candeloroFeedbackAssistedQuantumSearch2023,Hutin2025}—which is the central objective of this work—but also for enhancing desirable quantum properties such as entanglement and squeezing~\cite{wisemanQuantumTheoryOptical1993,thomsenSpinSqueezingQuantum2002,serafiniDeterminationMaximalGaussian2010,genoniQuantumCoolingSqueezing2015,brunelliConditionalDynamicsOptomechanical2019,digiovanniUnconditionalMechanicalSqueezing2021,rossiPRL2020,AmorosBinefa2021,Fallani2022,isaksenMechanicalCoolingSqueezing2023,caprotti2024analysis,amorosbinefa2024}, and, more recently, to optimize quantum thermodynamics protocols~\cite{ElouardPRL2017,bhandariContinuousMeasurementBoosted2022,Morrone2023}. Experimentally, feedback-based state engineering has been successfully demonstrated across various platforms, including circuit-QED and quantum optomechanical systems~\cite{sayrin2011real,vijay2012stabilizing,riste2013deterministic,cox2016deterministic,liu2016comparing,campagne2016using,rossi2018control,magriniRealtimeOptimalQuantum2021,tebbenjohannsQuantumControlNanoparticle2021,navid2025experimental,dassonneville2025directlyprobingworkextraction}. 

While optimal control strategies can be analytically determined in specific cases, finding an optimal feedback strategy in general is a complex challenge. This difficulty is underscored by recent proposals leveraging deep reinforcement learning (DRL) techniques~\cite{Krenn2023,stevetutorial2025,ma2025} to address quantum state engineering (see for example~\cite{fosel2018reinforcement,Borah2021,porotti2022deep,Fallani2022,Hutin2025}). Motivated by this, we aim to answer the question: {\it What is the maximum fidelity that can be achieved with a given target state using these protocols?} Answering this question not only provides a benchmark for assessing the performance of existing feedback strategies—whether derived analytically or numerically—but also offers insight into the fundamental controllability limits of a given quantum dynamics. 

Inspired by quantum speed limits~\cite{Deffner2017}, we derive an upper bound on the fidelity rate that applies universally to feedback-based state engineering protocols. This leads to a tighter upper bound on the steady-state fidelity, improving upon the result previously established by Kobayashi and Yamamoto~\cite{kobayashi2019control}. Crucially, calculating this bound does not require any simulation or optimization, it can be calculated using only the Lindblad jump operators, Hamiltonian variance, and target state. Beyond presenting the general properties of the bound, we also specialize it to the case of Markovian feedback~\cite{Wiseman1994}, analyzing its dependence on the time resolution of the feedback strategy. 

We then apply our findings to the problem of preparing Dicke states~\cite{Dicke1954} in systems subject to collective damping and dispersive coupling. Dicke states naturally arise in many experimental platforms, particularly in ensembles of $N$ two-level atoms. Moreover, they serve as important resources for quantum information processing. For example, the highly entangled Dicke state $\ket{l=N/2,m=0}$ is valuable for quantum metrology~\cite{Hyllus2012}, and is known to exhibit robustness against particle loss~\cite{stockton2003characterizing,Guhne2008}. The Dicke state $\ket{l=N/2,m=N/2}$ corresponds to the maximally excited state and is relevant for quantum battery applications~\cite{CampaioliRMP2024}. For this control problem, we derive two additional, tighter bounds by imposing constraints on the control Hamiltonian and the monitoring strategies. Finally, we propose control protocols designed to prepare the aforementioned specific Dicke states, benchmarking them against our fidelity bounds and further validating their optimality by comparing them to numerical strategies obtained using DRL algorithms.
 
The manuscript is structured as follows: in Sect.~\ref{Sec:Deriv} we introduce the quantum state engineering problem and provide a step-by-step derivation of an upper bound for the steady-state fidelity between a given target pure state and the state obtained via a generic quantum control protocol based on continuous monitoring and feedback. In particular, Sec.~\ref{s:markovianFB} discusses modifications to this bound when restricting to Markovian feedback strategies. In Sec.~\ref{Sec:Dicke} we apply our fidelity bound to the problem of preparing Dicke states under collective damping and dispersive coupling, deriving large $N$ scaling results. In Sec.~\ref{Sec:class_lim} we place additional contrictions on the control Hamiltonian and derive two additional, tighter bounds. Sec.~\ref{Sec:opt_cont} focuses on the optimal control problem: inspired by the inequalities used in deriving our bounds, we propose specific control strategies, compare their steady-state fidelities against our upper bounds, and numerically validate their optimality by benchmarking them against DRL-based strategies. Finally, Sec.~\ref{sec:conclusions} concludes the paper with final remarks and future outlooks.

\section{Bounding the average fidelity for feedback control protocols}
\label{Sec:Deriv}
\subsection{Setting the scene: measurement-based feedback control protocols}
\label{Sec:measure}
We are interested in a quantum system whose dynamics is described by a Markovian master equation in the Lindblad form
\begin{equation}
\frac{d\rho(t)}{dt} = - i[\hat{H}, \rho]   +\sum_j \mathcal{D}[\hat{c}_j]\rho \label{eq:ME}
\end{equation}
where $\mathcal{D}[\hat{A}]\rho = \hat{A}\rho\hat{A}^\dag - (\hat{A}^\dag \hat{A} \rho + \rho \hat{A}^\dag \hat{A})/2$ is the standard Lindblad dissipator, and $\hat{H}$ is the Hamiltonian that we will discuss later in more detail.
Each operator $\hat{c}_j$ identifies a noisy channel that can be potentially monitored continuously in time. We now in fact assume that continuous homodyne detection is performed on the corresponding output channels, leading to a conditional evolution for the quantum states $\rho^{(c)}$, which are typically referred to as quantum trajectories~\cite{wisemanQuantumMeasurementControl2009,albarelli2024pedagogical}. This evolution is described by a 
diffusive stochastic master equation of the form
\begin{align}
    d\rho^{(c)}(t) &= - i[\hat{H}^{(c)}, \rho^{(c)}] dt  +\sum_j \mathcal{D}[\hat{c}_j]\rho^{(c)} dt  \nonumber \\
   &\,\,\,\,\,  + \sum_j \sqrt{\eta_j}\mathcal{H}[\hat{c}_j]\rho^{(c)} dw_j\,, \label{eq:SME}
\end{align}
conditioned on the corresponding homodyne photocurrents, whose increments read
\begin{equation}
dy_j = \sqrt{\eta_j} \tr[\rho(\hat{c}_j +\hat{c}_j^\dag)] dt + dw_j \,.\label{eq:photocurrents}
\end{equation}  
In Eq.~(\ref{eq:SME}) we have introduced the non-linear superoperator $\mathcal{H}[\hat{A}]\rho = \hat{A}\rho + \rho \hat{A}^\dag - \tr[\rho(\hat{A}+\hat{A}^\dag)]\rho$, that represents the update in our knowledge of the system state based on the measurement result, while the parameters $\eta_j$ quantify the measurement efficiencies of each homodyne detection. The quantities $\{dw_j\}$, which connect the measured photocurrents in (\ref{eq:photocurrents}) and the stochastic quantum evolution in (\ref{eq:SME}), mathematically correspond to uncorrelated stochastic Wiener increments, satisfying the Ito's lemma formula $dw_j dw_k = \delta_{jk} dt$, and whose stochastic average is zero, i.e. $\mathbbm{E}[dw_j] =0$. Utilizing this last property one can easily prove that by considering $\mathbbm{E}[d\rho^{(c)}]$, that is averaging the conditional evolution over all the possible measurement outcomes and thus over all the possible quantum trajectories, one obtains the unconditional master equation (\ref{eq:ME}).

The Hamiltonian $\hat{H}^{(c)} = \hat{H}_0 + \hat{H}_u^{(c)}(t)$ consists of a bare Hamiltonian, $\hat{H}_0$ and a control Hamiltonian $\hat{H}_u^{(c)}(t)$. We will consider the so-called {\it state-based control} scenario, that is, while we denote a time-dependence of the control Hamiltonian, it will actually depend on the measurement results obtained up to time $t$, and thus in the best-case scenario, on the conditional state of the system $\rho^{(c)}$ at time $t$.
To simplify the notation, we will use an overbar to denote trajectory-averaged quantities, e.g. $\overline{\rho}=\mathbbm{E}[\rho^{(c)}]$.

We would like to emphasize that once a Lindblad master equation of the form in Eq. (1) is given—possibly with an explicitly time-dependent Hamiltonian $\hat{H}^{(c)}(t)$ as defined above—it always defines a completely positive and trace-preserving (CPTP) dynamical map. This guarantees that the evolution is physically valid for any such time-dependent Hamiltonian. Accordingly, from a purely fundamental point of view, the analysis we will provide in this manuscript is applicable to any dynamics governed by such a master equation, regardless of the specific time dependence in $\hat{H}^{(c)}(t)$. It is however true that adding a time-dependent Hamiltonian into a Lindblad equation risks violating the assumptions under which the Lindblad form was originally derived: this can be done as long as one can consider the bath as delta-correlated~\cite{deVega2010}, thus implying that the time-dependence in the Hamiltonian must evolve on a timescale slower than the bath correlation time. For this reason, given a certain experimental scenario and a given control strategy, one will have to carefully address this potential issue, for example by smoothing the corresponding control protocol to ensure consistency with Lindblad dynamics.

The fidelity, $\F = \ev{\rho^{(c)}}{\psi_T}$, to a pure target state is a ubiquitous measure of success in quantum control problems. Physically, it quantifies the probability that the system will be found in the target state when measured in a suitable basis. In the presence of continuous monitoring, the fidelity becomes a stochastic quantity.
We will focus on bounding the average fidelity $\overline{\mathcal{F}} = \mathbbm{E}[\!\ev{\rho^{(c)}}{\psi_T}]=\ev{\overline{\rho}}{\psi_T}$. On the one hand this quantity quantifies on average how good is our feedback-control quantum state engineering protocol, in the scenario where one has knowledge of the measurement outcomes obtained during the continuous monitoring, and thus on the particular quantum trajectory $\rho^{(c)}$ prepared in each single run of the experiment. On the other hand, because of the linearity of the average, we have $\overline{\F} =\langle\psi|\overline{\rho}|\psi\rangle$, that is, it also corresponds to the fidelity between the target state $|\psi\rangle$ and the average state $\overline{\rho}=\mathbb{E}[\rho^{(c)}]$; this observation thus extends the operational meaning of $\overline{\F}$ to the scenario where the measurement apparatus and the feedback controller operate in a {\it black-box}, and thus one has no knowledge of the measurement outcomes, and of each single quantum trajectory $\rho^{(c)}$; the output of such protocol is indeed fully described by the unconditional state $\overline{\rho}$ obtained by averaging over all the possible measurement outcomes. To improve readability, $|\psi_T\rangle$ will be denoted simply as $|\psi\rangle$ throughout the manuscript, and this symbol will exclusively refer to the target state.

An obvious place to start is to look at the behavior of the fidelity with time, by considering its stochastic increment $d\mathcal{F} = \ev{d\rho^{(c)}}{\psi}$, which can be written as
\begin{align}
  d\mathcal{F} = \langle\mathscr{U}\rho^{(c)}\rangle_{\psi} dt +  \langle \mathscr{D} \rho^{(c)} \rangle_{\psi} dt + \langle \mathscr{H} \rho^{(c)} \rangle_{\psi,\{dw_j\}} \,.
\end{align}
Here we have defined the unitary term
\begin{align}
 \langle\mathscr{U} \rho^{(c)}\rangle_{\psi} =     -i\ev{[\hat{H}^{(c)}, \rho^{(c)}]}{\psi} \,, \label{eq:unitaryterm}
\end{align}
the noisy term
\begin{align}
 \langle\mathscr{D}\rho^{(c)}\rangle_{\psi} =   \sum_j \bra{\psi}\mathcal{D}[\hat{c}_j]\rho^{(c)}\ket{\psi}, \label{eq:noisyterm}
\end{align}
and the stochastic term
\begin{align}
 \langle\mathscr{H}\rho^{(c)}\rangle_{\psi,\{dw_j\}} =   \sum_j \bra{\psi}\mathcal{H}[\hat{c}_j]\rho^{(c)}\ket{\psi}\,dw_j \,.
 \label{eq:stochasticterm}
\end{align}
We start by considering $\langle\mathscr{H}\rho^{(c)}\rangle_{\psi,\{dw_j\}}$. 
As we will be interested in the average fidelity $\overline{\mathcal{F}}$ and thus on its average increment $\overline{d\mathcal{F}}$, we take its stochastic average and observe
that, 
\begin{align}
\overline{\langle\mathscr{H}\rho\rangle}_{\psi,\{dw_j\}} &= \mathbbm{E}\left[ \sum_j \bra{\psi}\mathcal{H}[\hat{c}_j]\rho^{(c)}\ket{\psi}\,dw_j \right] \nonumber \\
&=  \mathbbm{E}\left[ \sum_j \bra{\psi}\mathcal{H}[\hat{c}_j]\rho^{(c)}\ket{\psi}\right] \mathbbm{E}\left[\,dw_j \right] \nonumber = 0 \,,
\end{align}
where we have exploited the fact that the Wiener increments $dw_j$ and the conditional states $\rho^{(c)}$  are independent.
We have thus proven that the stochastic term (\ref{eq:stochasticterm}) is giving an average-zero contribution to the average fidelity increment and we will thus end up with the unitary term (\ref{eq:unitaryterm}) and the noisy term (\ref{eq:noisyterm}) only, obtaining
\begin{align}
\frac{\overline{d\F}}{dt} = \overline{\langle\mathscr{U}\rho\rangle}_{\psi}  +  \overline{\langle \mathscr{D} \rho \rangle}_{\psi}\,. 
\label{eq:avgfid_rate}
\end{align}
In the following, we separately derive upper bounds for both terms. 

We finally remark that these bounds hold for all possible unravellings of the Markovian master equation~\eqref{eq:ME}: one can easily show that ${\langle\mathscr{H}\rho^{(c)}\rangle}_{\psi,\{dw_j\}}$ can be always replaced by a different stochastic term that will still give no contribution on average. This clearly includes quantum jump unravellings due to continuous photodetection.
\subsection{Bounding the unitary term via the quantum speed limit} 
\label{s:unitary}
To derive a non-trivial bound on the fidelity, we must impose restrictions on the \textit{strength} of our Hamiltonian. Here we look to the quantum speed limit (QSL) literature to determine a suitable measure for comparing the strength of different Hamiltonians~\cite{Deffner2017}. Many different measures have been suggested but the simplest and most common is the Mandelstam-Tamm bound which puts a limit on the variance of the Hamiltonian, 
\begin{align}
(\Delta \hat{H}^{(c)})^2 = \ev{(\hat{H}^{(c)})^2}{\psi}-\ev{\hat{H}^{(c)}}{\psi}^2 \leq (\Delta E)^2. \label{eq:varianceH}
\end{align}
$\Delta E$ is a constant that restricts the maximum value of $\Delta \hat{H}^{(c)} $ (one can find more details on its physical and mathematical meaning in Sec.~\ref{sec:bound_interp}). The Mandelstam-Tamm bound~\cite{Mandelstam1945} can be derived by applying Robertson's uncertainty relation~\cite{robertson1929} to the Schrödinger equation, leading to
\begin{align}
    \label{eq:robertson}
     \langle\mathscr{U}\rho^{(c)}\rangle_{\psi} &=  -i\ev{[\hat{H}^{(c)}, \rho^{(c)}]}{\psi}  \nonumber \\
   &\leq \left| \!\ev{[\hat{H}^{(c)}, \rho^{(c)}]}{\psi} \right| \nonumber \\
    & \leq 
     2 \Delta \hat{H}^{(c)} \sqrt{\ev{{\rho^{(c)}}^2}{\psi}- \ev{\rho^{(c)}}{\psi}^2} \nonumber\\
    &\leq 2 \Delta E \sqrt{\mathcal{F} - \mathcal{F}^2}\,,
\end{align}
where in the last line we exploited the inequality $\ev{{\rho^{(c)}}^2}{\psi} \leq \ev{\rho^{(c)} }{\psi}=\mathcal{F}$. By taking the average over all the trajectories we find
\begin{align}
\overline{ \langle\mathscr{U}\rho\rangle}_{\psi}  &\leq  2  \Delta E \, \mathbbm{E}\left[\sqrt{\mathcal{F} - \mathcal{F}^2}\right]
\nonumber \\ 
 &\leq 2 \Delta E \textstyle\sqrt{\,\overline{\mathcal{F}} - \smash{\overline{\mathcal{F}}}^2}
\label{eq:upperbound_unitary}
\end{align}
we have applied Jensen's inequality to the concave function $g(\mathcal{F}) = \sqrt{\mathcal{F} - \mathcal{F}^2}$. The fact that $\Delta E$ is constant allows us to ignore it when we take the average over trajectories, as we have done above.
\subsection{Bounding the noisy term} 
We can now focus on bounding the noisy term in Eq.~(\ref{eq:noisyterm}).
Since we are deriving a bound, we can consider the best-case scenario where each trajectory is described by a pure conditional state $\rho^{(c)}=|\varphi^{(c)}\rangle\langle \varphi^{(c)}|$; this is obtained whenever one considers an initial pure state and unit-efficiency detection $\eta_j=1$ for all the homodyne measurements. 

We can write any pure quantum state in the form 
\begin{align}
\ket{\varphi^{(c)}} = \sqrt{\mathcal{F}} \ket{\psi} + e^{i \phi^{(c)}} \sqrt{1-\mathcal{F}}\ket{\psi^{\perp(c)}}.
\label{eq:purestate}
\end{align}
Substituting this into Eq.~\eqref{eq:noisyterm} the result is (the full expansion and calculation can be found in Appendix~\ref{App:Drift})
\begin{align}
\label{eq:noisyterm1}
\langle \mathcal{D} \rho \rangle_{\psi} 
& \leq -\mathcal{F} (\Delta \mathbf{\hat{c}})^2 + (1-\mathcal{F})\mathcal{A}^{(c)} + \mathcal{B}^{(c)}\sqrt{\mathcal{F} - \mathcal{F}^2}
\end{align}
where
\begin{align}
    \label{eq:Delta_c}
    (\Delta \mathbf{\hat{c}})^2 &= \sum_j (\Delta \hat{c}_j)^2 \nonumber\\
    &= \sum_j\left(\expval{\hat{c}_j^\dag\hat{c}_j}{\psi} - \left| \! \expval{\hat{c}_j}{\psi}\right|^2\right), \\
    \label{eq:A}
    \mathcal{A}^{(c)} &= \sum_j\lvert \! \mel*{\psi^{\perp(c)}}{\hat{c}_j^\dag}{\psi}\rvert^2,\\ 
    \label{eq:B}
    \mathcal{B}^{(c)} &= \bigg\lvert\bra*{\psi^{\perp(c)}}\sum_j \left(2\hat{c}_j^\dag \dyad{\psi} \hat{c}_j - \hat{c}_j^\dag\hat{c}_j\right) \ket*{\psi} \bigg\rvert.
\end{align}
$(\Delta \mathbf{\hat{c}})^2$ captures the rate at which the environment removes population from the target state. On the other hand, $\mathcal{A}^{(c)}$ governs the rate at which the environment brings population from the orthogonal subspace into the target state. $\mathcal{B}^{(c)}$ is an interference term that doesn't have such a clear interpretation. Before averaging over trajectories we must optimize the upper bound in Eq.~\eqref{eq:noisyterm1} over the orthogonal state $|\psi^{\perp(c)}\rangle$ that we introduced in Eq.~\eqref{eq:purestate}. In general, it is not possible to simultaneously maximize both $\mathcal{A}^{(c)}$ and $\mathcal{B}^{(c)}$ terms independent of $\mathcal{F}$. Therefore, we will maximize each individually, denoting them $\mathcal{A}^*=\max_{\psi^\perp}\mathcal{A}^{(c)}$ and $\mathcal{B}^*=\max_{\psi^\perp}\mathcal{B}^{(c)}$, which can be calculated given $\ket{\psi}$ and $\{c_j\}$ only. Thus we obtain quantities that do not depend on the conditional state $|\varphi^{(c)}\rangle$. By taking the average over trajectories, and by applying Jensen's inequality once again to the function $g(\mathcal{F})=\sqrt{\mathcal{F} - \mathcal{F}^2}$, we obtain an upper bound on the noisy term
\begin{align}
\label{eq:upperbound_noisy}
\overline{\langle \mathcal{D} \rho \rangle}_{\psi} 
& \leq -\overline{\mathcal{F}} (\Delta \mathbf{\hat{c}})^2 + (1-\overline{\mathcal{F}})\mathcal{A}^* + \mathcal{B}^*\textstyle\sqrt{\,\overline{\mathcal{F}} - \smash{\overline{\mathcal{F}}}^2}
\end{align}
$\mathcal{A}^*$ and $\mathcal{B}^*$ can be calculated explicitly and take the form
\begin{align}
    \mathcal{A}^* &= \lVert \hat{Q} \hat{X}_{\mathcal{A}} \hat{Q} \rVert_{\sf op} \\
    \mathcal{B}^* &= \Delta \hat{X}_{\mathcal{B}}
\end{align}
Where $\lVert \hat{O} \rVert_{\sf op}$ represents the operator norm of $\hat{O}$ and $\hat{Q} = \hat{\mathbbm{1}} - \dyad{\psi}$ is the projector onto the orthogonal subspace. $X_{\mathcal{A}}$ and $X_{\mathcal{B}}$ are defined as
\begin{align}
    \hat{X}_{\mathcal{A}} = \Bigg(\sum_j \hat{c}_j^\dag \dyad{\psi} \hat{c}_j \Bigg)\\
    \hat{X}_{\mathcal{B}} = \sum_j \left(2\hat{c}_j^\dag \dyad{\psi} \hat{c}_j - \hat{c}_j^\dag\hat{c}_j\right).
\end{align}
A full derivation can be found in Appendix \ref{App:Drift}.
%
%
%
\subsection{Deriving and interpreting the fidelity upper bound}
\label{sec:bound_interp}
We can now sub Eqs.~\eqref{eq:upperbound_unitary} and \eqref{eq:upperbound_noisy} into Eq.~\eqref{eq:avgfid_rate} and use the fact that $\overline{d \mathcal{F}} = d \overline{\mathcal{F}}$ to get an upper bound on the rate of change of the average fidelity
\begin{equation}
    \label{eq:fid_rate_sol}
    \frac{d \overline{\mathcal{F}}}{dt} \leq (\mathcal{B}^* + 2\Delta E) \textstyle\sqrt{\,\overline{\mathcal{F}} - \smash{\overline{\mathcal{F}}}^2} -\overline{\mathcal{F}} (\Delta \mathbf{\hat{c}})^2 + (1-\overline{\mathcal{F}})\mathcal{A}^*.
\end{equation}
This inequality provides a bound on the rate of change of the instantaneous average fidelity, $\overline{\mathcal{F}}(t)$, and thus takes the form of a QSL in the presence of continuous monitoring and state-based feedback, and we can thus refer to it as a QSL for feedback-based quantum state engineering protocols. It is worth noting that this QSL is still useful in cases where the dynamics considered allow unit fidelity at steady-state; in such scenarios, Eq.~\eqref{eq:fid_rate_sol} bounds the rate at which unit fidelity can be approached.
In general, given some initial state fidelity, we can integrate this inequality to get an upper bound on $\overline{\mathcal{F}}(\tau)$ at a later time, $\tau$. Alternatively, we can set $d\overline{\mathcal{F}}/dt = 0$ and solve Eq.~\eqref{eq:fid_rate_sol} to obtain an upper bound on the average steady-state fidelity, $\overline{\mathcal{F}}_{ss}$, independent of the initial state,
\begin{widetext}
\begin{equation}
    \label{eq:gen_sol}
  \overline{\mathcal{F}}_{ss}  \leq  \mathscr{B}_\text{QSL}   \equiv \frac{2\mathcal{A}^*(\mathcal{A}^* + (\Delta \mathbf{\hat{c}})^2) + (\mathcal{B}^* +2 \Delta E) \left((\mathcal{B}^* +2 \Delta E) + \sqrt{4\mathcal{A}^*(\Delta \mathbf{\hat{c}})^2 +(\mathcal{B}^* +2 \Delta E)^2} \right)}{2\left((\mathcal{A}^* + (\Delta \mathbf{\hat{c}})^2)^2 + (\mathcal{B}^* +2 \Delta E)^2\right)}.
\end{equation}
\end{widetext}
We will refer to this bound as the {\it QSL-bound}.
A simplified, but also illuminating form of the bound is obtained when $\mathcal{A}^*=\mathcal{B}^*=0$, yielding
\begin{equation}
    \label{eq:BQ_simple}
  \overline{\mathcal{F}}_{ss}  \leq  \mathscr{B}_\text{QSL}   
= \frac{1}{1+\left[(\Delta \mathbf{\hat{c}})^2/(2\Delta E)\right]^2} \,,
\end{equation}
where the competition between the environment and the Hamiltonian is clear. Notice however, that as we will see in the examples provided in the next section, this condition is not in general satisfied.

We remark how $\mathscr{B}_\text{QSL}$ depends only on the target state $|\psi\rangle$, the Hamiltonian \textit{strength} $\Delta E$, and the Lindblad operators $\{\hat{c}_j\}$.
As regards to its other properties, we observe that 
\begin{align}
\mathscr{B}_\text{QSL} = 1 \iff (\Delta \mathbf{\hat{c}})^2 = 0.
\label{eq:unitfidelity}
\end{align}
Given that each individual term in $(\Delta \mathbf{\hat{c}})^2$ is non-negative, $(\Delta \mathbf{\hat{c}})^2 = 0$ implies $\ket{\psi}$ is an eigenvector of all $\{\hat{c}_j\}$, and thus corresponds to a steady-state of the noisy dynamics~\cite{Kraus2008}. Finally we observe that the bound is monotonically increasing with $\Delta E$, $\mathcal{A}^*$, and $\mathcal{B}^*$, while it is monotonically decreasing with $(\Delta \mathbf{\hat{c}})^2$ (these properties can be understood by looking at the QSL in Eq. (\ref{eq:fid_rate_sol})).

It is now worth taking a moment to understand the kind of control strategies we allowed in deriving the upper bound $\mathscr{B}_\text{QSL}$. As we explained in Sec.~\ref{s:unitary}, restrictions must be placed on the Hamiltonian but determining exactly what those restrictions should be is not always obvious. This difficulty is perhaps most clearly exemplified in the zoo of different quantum speed limits that have been derived~\cite{Margolus1998, Taddei2013, Deffner2013Open, DelCampo2013, Pires2016, ModiPRL, garcia2019quantum, oconnor2021action}. There is always a trade-off between the tightness of a bound for a specific problem and the applicability of that bound to a large class of problems. The bound $\mathscr{B}_\text{QSL}$ allows for significant freedom in the choice of control Hamiltonian: it holds for any Hamiltonian $\hat{H}^{(c)}(t,\rho)$ that satisfies 
\begin{equation}
    \label{eq:control_strength}
    \Delta \hat{H}^{(c)}(t,\rho) \leq \Delta E \quad \forall \, t,\rho,
\end{equation}
for a given maximum strength $\Delta E$. The first thing to note is that a restriction on the variance is very different from a restriction of the form $\norm{\hat{H}^{(c)}(t,\rho)} \leq \norm{E}$, because the variance \eqref{eq:varianceH} is taken with respect to the target state. Eq.~\eqref{eq:control_strength} does, in fact, allow for arbitrarily large control strength, but only within the subspace of states orthogonal to the target state. This has both a positive side, because we allow for a larger class of Hamiltonians, and a negative side, since, if needed, such high-strength Hamiltonians may be unrealistic. When the target state is an eigenstate of the bare Hamiltonian, as is very common in control problems, the variance, unlike the norm, does not depend on the bare Hamiltonian, that is $\Delta \hat{H}^{(c)} = \Delta \hat{H}_u^{(c)}$.

Another advantage of using $\Delta E$ as a constraint, is that the Hamiltonian that saturates Eq.~\eqref{eq:robertson} is known and for a pure state $\ket{\varphi^{(c)}}$, decomposed as in Eq.~\eqref{eq:purestate}, it takes the form~\cite{Brody2006}
\begin{equation}
    \label{eq:QSL_Ham}
    \hat{H}^{(c)}(t,\ket{\varphi^{(c)}}) = \frac{\Delta E}{\sqrt{2}} \left(e^{i (\phi - \pi/2)}\dyad*{\psi}{\psi^\perp} + e^{-i (\phi - \pi/2)}\dyad*{\psi^\perp}{\psi}\right).
\end{equation}
When $\mathcal{A}$ and $\mathcal{B}$ are maximizable by the same orthogonal state, $\ket*{\psi^\perp}$, we can immediately derive a control strategy that saturates Eq.~\eqref{eq:fid_rate_sol} (up to Jensen's inequality): we apply infinite control strength in the orthogonal subspace to ensure that $\ket*{\psi^\perp}$ is the optimal orthogonal state and simultaneously we apply the optimal Hamiltonian from Eq.~\eqref{eq:QSL_Ham}. This tells us that, in practice, the bound will be tightest when $\ket*{\psi^\perp}^*$ is easy to stabilize and our Hamiltonian can transfer population from $\ket*{\psi^\perp}^*$ to $\ket{\psi}$ at the quantum speed limit.

As we anticipated in the introduction, Kobayashi and Yamamoto~\cite{kobayashi2019control} derived a similar bound under the same assumptions, we label that bound $\mathscr{B}_\text{KY}$. In Appendix~\ref{app:KYBound} we prove that our bound is always tighter, that is $\mathscr{B}_\text{QSL} \leq \mathscr{B}_\text{KY}$ and we also show that there is a non-negligible gap between the two bounds for the examples (target states and dynamics) we are going to consider in the next sections.

Another advantage of our bound with respect to the one derived in~\cite{kobayashi2019control} is that it directly highlights the optimal control strategy under the corresponding constraints.  As we will see in the next section, it is possible to derive tighter bounds for certain classes of problems; however, these bounds involve placing additional constraints on the control strategy and the interaction with the environment. 

\subsection{Deriving the upper bound for Markovian feedback}\label{s:markovianFB}
In state-based feedback, we can condition our control Hamiltonian on the full measurement record up until time $t$. This is more general than Markovian feedback where the control Hamiltonian depends only on the most recent measurement result~\cite{Wiseman1994}. However, Markovian feedback protocols have been proposed that can stabilize the target state even when $(\Delta \mathbf{\hat{c}})^2 \neq 0$ \cite{hofmann1998quantum,wang2001feedback,ticozzi2009analysis,campagne2016using}. In this section, we will clarify this apparent contradiction and derive a bound on state stabilization for Markovian feedback. We will consider Markovian feedback of the form 
\begin{equation}
    \hat{H}^{(c)} = H_0 + \hat{F} I(t),
    \label{eq:Mark_Ham}
\end{equation}
where $\hat{F}$ is fixed and $I(t) = dy/dt$ is a homodyne photocurrent described by Eq.~\eqref{eq:photocurrents}. For simplicity we will assume that $\ket{\psi}$ is an eigenstate of $H_0$ and therefore does not contribute to $\Delta \hat{H}^{(c)}$. It becomes clear that this feedback strategy violates our bounds since $\mathbbm{E}[|I(t)|] \sim \frac{1}{\sqrt{dt}}$ is unbounded. In practice, the measurement current must be recorded over a finite time interval $\Delta t$, the measured signal will then be
\begin{equation}
    \Delta y_t = \int_{t-\Delta t}^{t}dy_{t'} = \int_{t-\Delta t}^{t}dt' \sqrt{\eta} \tr[\rho^{(c)}(\hat{c} +\hat{c}^\dag)] + \int_{t-\Delta t}^{t} dw.
    \label{eq:finite_increment}
\end{equation}
Under this assumption, we prove in Appendix~\ref{app:mark_fb} that it is not possible to stabilize target states with $(\Delta \mathbf{\hat{c}})^2 \neq 0$. We derive an equivalent bound to Eq.~\eqref{eq:fid_rate_sol} with
\begin{equation}
    \Delta E \to \left(\lVert\hat{c} +\hat{c}^\dag\rVert_{\sf op}  + \frac{1}{\sqrt{\Delta t}}\right) \Delta \hat{F}\,,
\end{equation}
where $\lVert \hat{O} \rVert_{\sf op}$ denotes the operator norm.
Additionally, we show that when $\lVert\hat{c} +\hat{c}^\dag\rVert_{\sf op}$ is finite, stabilizing a state with $1-\mathcal{F}_{ss} \leq \varepsilon$ for small $\varepsilon$ requires a vanishing measurement window
\begin{equation}
    \Delta t \leq \varepsilon \left(\frac{2 \Delta \hat{F}}{(\Delta \mathbf{\hat{c}})^2}\right)^2,
\end{equation}
and a diverging average Hamiltonian strength
\begin{equation}
    \mathbbm{E}[\Delta \hat{H}^{(c)}] \geq \sqrt{\frac{2}{\pi \varepsilon}}(\Delta \mathbf{\hat{c}})^2.
\end{equation} 
We confirm this scaling behavior by analyzing the stabilization of a qubit in the excited state, in the scenario where spontaneous emission can be continuously monitored~\cite{campagne2016using} (see Appendix~\ref{app:mark_fb} for details).
\section{Applying the bound to Dicke state preparation}
\label{Sec:Dicke}
We consider an ensemble of $N$ two-level atoms, described by collective spin operators $\hat{J}_i = \sum_n \hat{\sigma}_{i,n}/2$, with $i=\{x,y,z\}$ and where $\hat{\sigma}_{i,n}$ denotes the $i$-th Pauli operator of the $n$-th atom. In particular, we will be interested in the preparation of Dicke states $\ket{l,m}$, that is the common eigenstates of
\begin{align}
    \hat{\mathbf{J}}^2\ket{l,m} &= l(l+1)\ket{l,m}\\
    \hat{J}_z\ket{l,m} &= m \ket{l,m},
\end{align}
where $\hat{\mathbf{J}}^2 = \hat{J}_x^2 + \hat{J}_y^2 + \hat{J}_z^2$. We will assume that the unmonitored evolution is described by a Lindblad master equation of the form
\begin{align}
\frac{d\rho}{dt} = -i [\hat{H}^{(c)}(t), \rho] + \kappa \mathcal{D}[\hat{J}_z]\rho + \gamma \mathcal{D}[\hat{J}_-]\rho \label{eq:MEDicke} \,,
\end{align}
where $\hat{J}_- =  \sum_n \hat{\sigma}_{-,n}$, with $\hat{\sigma}_- = (\hat{\sigma}_x - i \hat{\sigma}_y)/2$. From a physical point of view, the first dissipative channel, formally described by the jump operator $\hat{c}_1 = \sqrt{\kappa}\hat{J}_z$, corresponds to a quantum non-demolition (QND) interaction that can be obtained by dispersively coupling the atoms to a cavity mode, allowing the QND monitoring of the operator $\hat{J}_z$~\cite{Doherty1999,thomsenSpinSqueezingQuantum2002,Geremia2003,Molmer2004,Albarelli_2017,caprotti2024analysis,Barberena2024}; the second channel, described by the jump operator $\hat{c}_2 = \sqrt{\gamma} \hat{J}_-$, corresponds to a collective damping channel, and approximately describes a two-level systems uniformly coupled to a single, lossy electromagnetic field mode (in the case of a single atom, $N=1$, it describes its spontaneous emission). As for the dispersive coupling, in order to experimentally observe such dynamics, one can place the emitters in a cavity where they share  the same electromagnetic mode~\cite{RitschRMP2013,Norcia2016}, but it has been recently realized also in free space~\cite{Ferioli2023,Goncalves2025}. Remarkably such dynamics allows to observe a non-equilibrium superradiant phase transition in free space~\cite{carmichael1977,agarwal1977,morrison2008,Hannukainen2018,Ferioli2023,Goncalves2025} corresponding to a boundary time-crystal~\cite{iemini2018,montenegro2023}. In what follows, we will assume that both environmental channels can be continuously monitored. Additionally, we apply a state-based control Hamiltonian $\hat{H}_u^{(c)}(t)$ (in what follows we can safely consider no bare Hamiltonian $\hat{H}_0=0$, as it will be removed going into the interaction picture respect to the free Hamiltonian of the atoms). We consider two possible control Hamiltonian operators; we will refer to the first as the {\it optimized-phase} (OP) control Hamiltonian,
\begin{equation}
    \label{eq:OPHamiltonian}
    \hat{H}_{OP}^{(c)}(t) = u(t)\left(\cos\theta(t)\,\hat{J}_x + \sin\theta(t)\, \hat{J}_y\right) \,,
\end{equation}
where we allow optimization over not only the Hamiltonian coupling parameter $u(t)$, but also a phase parameter $\theta(t)$; we will, in turn, refer to the second as the {\it fixed-phase} (FP) control Hamiltonian, which reads
\begin{equation}
    \label{eq:FPHamiltonian}
    \hat{H}_{FP}^{(c)}(t) = u(t)\hat{J}_y \,, 
\end{equation}
where one can only optimize the strength $u(t)$, while allowing rotations only around the $y$-axis of the spin. We remind that, as we are dealing with state-based feedback control, in the Hamiltonians above the strength $u(t)$ and $\theta(t)$ are optimized given the full knowledge of the conditional state $\rho^{(c)}$ which is continuously updated during the dynamics, given the full photocurrents outputs.
The maximum variance $\Delta E$ for a given target Dicke state $\ket{\psi}=\ket{l_T,m_T}$ is the same for both Hamiltonians, yielding
\begin{equation}
    \label{eq:Dicke_var}
    \Delta E = \frac{\Tilde{u}}{\sqrt{2}}\sqrt{l_T(l_T+1) - m_T^2}\,,
\end{equation}
where $\Tilde{u} \geq \lvert u(t) \rvert $ is an upper bound on the control Hamiltonian coupling $u(t)$. We remind here that, as we discussed in Sec.~\ref{sec:bound_interp}, in terms of the bound $\mathscr{B}_\text{QSL}$ the constraint above still allows infinite-strength control Hamiltonians that have support orthogonal to the target state. As we will see in the next sections, this assumption will be important to interpret the bound's achievability and to derive bounds tighter than $\mathscr{B}_\text{QSL}$.

Let us now calculate the terms in Eqs~\eqref{eq:Delta_c},~\eqref{eq:A} and \eqref{eq:B}. By exploiting the definition of Dicke states, in particular the fact that they are eigenstates of $\hat{J}_z$, we immediately obtain that the jump operator $\hat{c}_1 = \sqrt{\kappa}\hat{J}_z$ does not give any contribution to $\overline{\langle \mathcal{D} \rho \rangle}_{\psi}$. The jump operator $\hat{c}_2=\sqrt{\gamma} \hat{J}_-$,  satisfies the following property,
\begin{align}
\hat{J}_- |l,m\rangle= h_{m-1}\ket{l_T,m_T\!-\!1}.
\label{eq:collective_damping}
\end{align} 
where $h_m = \sqrt{(l-m)(l+m+1)}$. It follows that
\begin{align}
    (\Delta \mathbf{\hat{c}})^2 &= \gamma h_{m_T-1}^2\,, \\
    \mathcal{B}^* &= \mathcal{B} =0  \,.
\end{align}
Regarding the term
\begin{equation}
    \mathcal{A}^{(c)} = \gamma\lvert\bra*{\psi^\perp} \hat{J}_+\ket*{l_T,m_T}\rvert^2 \,,
\end{equation}
clearly the optimal orthogonal state is $\ket*{\psi^\perp}=\ket{l_T,m_T\!+\!1}$ as it is the only Dicke state that makes any contribution. We thus obtain
\begin{equation}
    \mathcal{A}^* = \gamma h_{m_T}^2.
\end{equation}
One can then sub these results into Eq.~\eqref{eq:gen_sol} and obtain an expression for the bound $\mathscr{B}_\text{QSL}$ for a generic Dicke state; as the corresponding general formula is cumbersome and not particularly informative, we will provide such results only for specific and relevant target states in the next section of the manuscript. We can, however, discuss some of the main properties of this bound: as we highlighted above the $\hat{J}_z$ channel does not give any contribution and thus the bound does not depend on the dispersive coupling $\kappa$; as a consequence, for collective damping, $\gamma=0$, one finds $\mathscr{B}_\text{QSL}=1$, thus placing no restriction on the steady-state fidelity $\overline{\mathcal{F}}_{ss}$. This aligns with previous results presented in~\cite{mirrahimi2007stabilizing,stockton2004deterministic}, where multiple control strategies have been derived that can deterministically prepare at steady-state any Dicke state using homodyne detection on the $\hat{J}_z$ channel and state-based feedback.


%
%
\subsection{Special cases}
We will now calculate $\mathscr{B}_\text{QSL}$ for two specific Dicke states and discuss the results. 
We start by considering the Dicke state $\ket{\psi}=\ket{N/2,N/2}$, corresponding to the maximally excited state ($l_T=m_T=N/2$).
The corresponding bound becomes
\begin{equation}
    \label{eq:ll}
    \overline{\mathcal{F}_{ss}} \leq \mathscr{B}_\text{QSL} = \frac{1}{1+ N \left(\frac{\gamma}{\Tilde{u}}\right)^2}.
\end{equation}
Here, $\mathcal{A}^*=0$, this tells us that no infinite-strength Hamiltonian is required, making the bound more likely to be accurate. As expected, the bound is decreasing as a function of the ratio $\gamma/\Tilde{u}$.
We also observe that if we consider the large $N$ limit, at fixed values of $\gamma$ and $\Tilde{u}$, the maximum achievable fidelity will tend to zero, meaning the system is completely uncontrollable.

The second case we will consider is the highly entangled Dicke state, $\ket{N/2,0}$,  i.e. $l_T=N/2$ for even $N$ and $m_T = 0$. The corresponding bound becomes
\begin{align}
    \label{eq:l0}
    \overline{\mathcal{F}_{ss}} & \leq \mathscr{B}_\text{QSL} = \frac12 + \frac{1}{\sqrt{4 + 2 N(N+2)\left(\frac{\gamma}{\Tilde{u}}\right)^2}} \,.
\end{align}
As in the previous case, we observe the expected monotonous behavior as a function of the ratio $\gamma/\Tilde{u}$. Moreover, in the large $N$ limit, it results in a bound for the fidelity of $\overline{\mathcal{F}} \leq 1/2$, hinting at the fact that $\ket{N/2,0}$ may be easier to prepare than $\ket{N/2,N/2}$, and could be engineered with a non-zero average fidelity in the large $N$ limit. However, as we discussed in Sec.~\ref{sec:bound_interp}, to achieve this bound it is necessary to stabilize the orthogonal state in $\ket{N/2,1}$ using an infinite-strength Hamiltonian (we remind that this kind of Hamiltonian would satisfy the constraint posed by Eq.~\eqref{eq:upperbound_unitary}). We will see in the next section that when one restricts the control Hamiltonians to \eqref{eq:OPHamiltonian} and \eqref{eq:FPHamiltonian} alone, $\ket{N/2,0}$ cannot be stabilized when $N \gg 1$.
\section{Tighter bounds for generation of Dicke states via monitoring and Feedback}
\label{Sec:class_lim}
\subsection{Restricting the allowed Hamiltonians}
\label{Sec:BD}
As mentioned earlier, one signature feature of $\mathscr{B}_\text{QSL}$ is that it allows for a huge family of control Hamiltonians. Of course, in practice, we are often restricted to a small family of control Hamiltonians. Unless this family happens to line up with the optimal strategy proposed in Sec.~\ref{sec:bound_interp} it is unlikely that $\mathscr{B}_\text{QSL}$ will be close to achievable. In this case, it would be valuable to derive a new bound that takes the features of the allowed control Hamiltonians into account. The Hamiltonian operators in Eqs.~\eqref{eq:OPHamiltonian} and \eqref{eq:FPHamiltonian} have an important property, population transfer can only occur between Dicke states with neighboring values of $m$. Additionally, the master equation terms corresponding to the noisy evolution in Eq.~\eqref{eq:MEDicke} also generate transfer solely between neighboring states. Similarly, the evolution preserves the value of $l$; for this reason, we will always assume that the initial state of the system belongs to the same subspace of the target state, i.e. it is a superposition of Dicke states with $l=l_T$. This assumption is particularly reasonable for $l_T=N/2$ as one can typically assume that the initial state is prepared in a coherent spin state, that is, a state obtained via a rotation around the Bloch sphere of the Dicke state $\ket{N/2,-N/2}$, and that corresponds to a product state where all the atoms have spin aligned along the same direction.

Under this assumption we can write the conditional pure state $|\varphi^{(c)}(t)\rangle$ at time $t$ in the Dicke basis as
\begin{align}
\ket{\varphi^{(c)}(t)} = \sum_{m=-l_T}^{l_T} e^{i \phi_m(t)}\sqrt{a_m(t)}\ket{l_T,m} \,,
\label{eq:pure_decomp}
\end{align}
where $\{a_m(t)\}$ and $\{\phi_m(t)\}$ are real numbers that correspond respectively to the population of the state $\ket{l_T,m}$ and to the corresponding relative phase parameter.
By plugging this conditional state into the most general stochastic master equation \eqref{eq:SME} corresponding to an unraveling of Eq.~\eqref{eq:MEDicke}, one can obtain a stochastic differential equation for the populations that will have the form
\begin{align}
da_m(t) &= T_{m} dt - T_{m-1}dt + d\mathcal{S}_m \,,
\label{eq:cse}
\end{align}
where $T_m$ denotes the deterministic transfer rate into the level $m$ from the level $m+1$, while $d\mathcal{S}_m$ denotes a sum of stochastic terms which are all linearly proportional to the Wiener increments that appear in Eq.~\eqref{eq:SME} to describe the backaction due to homodyne continuous monitoring.
When we employ the optimized-phase Hamiltonian $\hat{H}_{OP}$ in Eq.~\eqref{eq:OPHamiltonian}, the transfer rate has the form (see Appendix~\ref{app:transferrates} for more details on their derivation)
\begin{equation}
    \label{eq:transf_rates}
    T_m = u \sqrt{a_m a_{m+1}} \cos(\alpha_m) h_m  + \gamma a_{m+1} h_m^2 
\end{equation}
where we have defined $\alpha_m = \theta(t) -(\phi_m - \phi_{m+1}) - \pi/2$. 
Assuming the existence of a steady-state for the unconditional dynamics, in the long time limit $\mathbbm{E}[da_{m}] = 0$ for all values of $m$. Since $\mathbbm{E}[da_{-l}]=\mathbbm{E}[T_{-l}]\,dt=0$, following a cascade reasoning we obtain $\mathbbm{E}[T_{m}] = 0$ for all $m$. Explicitly
\begin{align}
    0&=\mathbbm{E}[T_m] \, \nonumber \\
      &= \mathbbm{E}[u \sqrt{a_m a_{m+1}} \cos(\alpha_m) h_m] + \gamma \overline{a}_{m+1} h_m^2 \,
\end{align}
For now we will focus on the states with $m \leq m_T$, to maximize the transfer rate \textit{towards} the target state we must minimize $T_m$. $T_m$ is minimized when $u \cos(\alpha_m) = -\tilde{u}$, it follows that
\begin{align}
    0 &\geq -\tilde{u} h_m\mathbbm{E}[\sqrt{a_m a_{m+1}}] + \gamma \overline{a}_{m+1} h_m^2 \, \nonumber \\
    & \geq -\tilde{u} h_m \sqrt{\overline{a}_m \overline{a}_{m+1}} + \gamma \overline{a}_{m+1} h_m^2 \,.
    \label{eq:transf_ineq}
\end{align}
In the final line, we have applied the Cauchy-Schwarz inequality. Rearranging, we can solve for $\overline{a}_m$:
\begin{align}
    \overline{a}_m &\geq \overline{a}_{m+1}\left(\frac{h_m \gamma}{\tilde{u}}\right)^2 \, \nonumber \\
    &\geq \overline{a}_{m_T}\prod_{k = m}^{m_T-1} \left(\frac{h_k \gamma}{\tilde{u}}\right)^2 \,.
    \label{eq:boundam}
\end{align}
The normalization of probabilities implies that
\begin{align}
    1 &= \sum_{m=-l_T}^{l_T}  \overline{a}_{m} \, \nonumber \\
    &\geq \sum_{m=-l_T}^{m_T}  \overline{a}_{m} \, \nonumber \\
    &\geq \overline{a}_{m_T} \left[1 + \sum_{m=-l_T}^{m_T-1} \prod_{k = m}^{m_T-1} \left(\frac{h_k \gamma}{\tilde{u}}\right)^2\right] \,.
    \label{eq:boundamT}
\end{align}
In the second line, we truncated the sum at $m=m_T$ and in the third line, we exploited the inequality in \eqref{eq:boundam}.
By reminding ourselves that $\overline{a}_{m_T} = \overline{\mathcal{F}}_{ss}$ and by rearranging Eq.~\eqref{eq:boundamT}, we obtain a bound for the steady-state fidelity with a generic Dicke state target 
\begin{align}
\overline{a}_{m_T} = \overline{\mathcal{F}}_{ss} \leq \mathscr{B}_\text{D} = \frac{1}{1 + \sum_{m=-l_T}^{m_T-1}\prod_{k = m}^{m_T-1} \left(\frac{h_k \gamma}{u}\right)^2}\,,
\label{eq:BD}
\end{align}
which we will refer to as the {\it Dicke-Bound}. Remarkably, the populations of the states with $m > m_T$ are irrelevant to the derivation of $\mathscr{B}_\text{D}$.

This bound is tighter than the one presented in the previous section, i.e. $\mathscr{B}_\text{D} \leq \mathscr{B}_\text{QSL}$, as it does not allow
infinite strength control Hamiltonians that change the Dicke state populations; moreover, it can be easily evaluated once $\gamma$ and $\tilde{u}$ are specified. 

Furthermore, we can see that when $l_T=N/2$ and $m_T>-N/2$, the sum in the denominator of $\mathscr{B}_\text{D}$ contains terms that scale as $(h_k)^2$. As a consequence, one obtains a vanishing average fidelity $\overline{\mathcal{F}}_{ss}$ in the limit of a large number of atoms, $N\gg 1$. We can thus conclude that with a Hamiltonian such as the ones in Eq.~\eqref{eq:OPHamiltonian} and~\eqref{eq:FPHamiltonian}, which are linear in the spin operators $\hat{J}_i$, one cannot prepare at steady-state any Dicke state $\ket{N/2,m_T}$ (except $m_T = -N/2$), unless $\gamma$ or $\Tilde{u}$ are functions of $N$. For this reason, we will choose to consider a dissipative parameter scaling inversely with the number of atoms, i.e. $\gamma = \tilde{\gamma}/N$ where now $\tilde{\gamma}$ is a fixed constant. Under this assumption, one can indeed obtain finite values for the average fidelity bound in the large $N$ limit. Remarkably, this choice is equivalent to considering a well-behaved master equation in the thermodynamic limit~\cite{Benatti2016,Benatti_2018,Carollo_2024}, showing a non-trivial connection between controllability and the existence of a well-defined thermodynamic limit of the open system dynamics. 

We remark that the same reasoning about the thermodynamic limit of the open system dynamics would also apply to the dispersive channel corresponding to the jump operator $\hat{c}_1 = \sqrt{\kappa}\hat{J}_z$. However, since $\mathscr{B}_\text{D}$ is independent of $\kappa$, we consider both constant $\kappa$ and inverse scaling $\kappa = \tilde{\kappa}/N$ and we will investigate the corresponding effect on the optimal control strategies in Sec.~\ref{Sec:opt_cont}.

\subsection{Specifying the monitoring strategy}
\label{Sec:BDS}
Despite additional restrictions on actual control Hamiltonian, the bound $\mathscr{B}_\text{D}$ is still very general. The bound still holds under the addition of arbitrary strength Hamiltonians that preserve the Dicke state populations $\{a_m\}$, enabling individual control of the relative phases $\{\phi_m\}$. Additionally, it encompasses all the possible unravellings of the master equations~\eqref{eq:ME}. We can now restrict ourselves to a particular monitoring strategy, one that is relevant both for experimental and theoretical reasons, and whose evolution is described by the following stochastic master equation
\begin{align}
    d\rho^{(c)}(t) &= \mathcal{L}\rho^{(c)} \,dt + \mathcal{H}[\sqrt{\kappa} \hat{J}_z]\rho^{(c)} dw_z  \nonumber \\
    &\,\, + \mathcal{H}[\sqrt{\gamma/2}\hat{J}_-]\rho^{(c)} dw_{x}  + \mathcal{H}[i\sqrt{\gamma/2} \hat{J}_-]\rho^{(c)} dw_{y}  \,.
     \label{eq:SME_Dicke}
\end{align}
The first term $\mathcal{L}\rho^{(c)}(t)\,dt$ denotes the unconditional evolution in Eq.~\eqref{eq:MEDicke}, the second term $\mathcal{H}[\sqrt{\kappa} \hat{J}_z]\rho^{(c)} dw_z $ describes the backaction due to the quantum non-demolition measurement of the operator $\hat{J}_z$, corresponding to the homodyne photocurrent
\begin{align}
dy_t^{(z)} = 2 \sqrt{\kappa} \Tr[\rho^{(c)}(t) \hat{J}_z] \,dt + dw_z \,,
\end{align}
while the last two terms describe the back-action due to the heterodyne measurement on the collective damping channel, corresponding to the double-homodyne photocurrents 
\begin{align}
dy_t^{(x)} &= \sqrt{2 \gamma} \Tr[\rho^{(c)}(t)  \hat{J}_x] \,dt + dw_x \,, \\
dy_t^{(y)} &= \sqrt{2 \gamma} \Tr[\rho^{(c)}(t)  \hat{J}_y] \,dt + dw_y \,.
\end{align}

Three inequalities were used to derive $\mathscr{B}_\text{D}$. The first one, exploited in Eq.~\eqref{eq:boundamT}, is  $\sum_{m=-l_T}^{m_T}  \overline{a}_{m} \leq 1$:  this implies that populating states with $m > m_T$ is not beneficial. In Eq.~\eqref{eq:transf_ineq} we used $u \cos(\alpha_m) \geq -\tilde{u}$: this tells us the optimal configuration of the relative phases $\phi_m$ for $m \leq m_T$,
\begin{equation}
    \Delta \phi_m \equiv \phi_m - \phi_{m+1} = \theta + \pi/2.
    \label{eq:phase_align}
\end{equation}
We thus conjecture that the optimal average steady-state fidelity is achieved when we saturate these two inequalities, and thus for states having the following form
\begin{equation}
    \ket{\varphi^{(c)}_{\sf opt}(t)} = \sum_{m=-l_T}^{m_T} e^{-i m \left(\theta + \frac{\pi}{2}\right)} \ket{l_T,m}
    \label{eq:opt_align}
\end{equation}

The third and final inequality applied is the Cauchy-Schwarz inequality. The tightness of this inequality is related to the distribution of $a_m$ and therefore depends sensitively on the control and measurement strategies implemented.

In Appendix~\ref{App:stochastic} we calculate $d\mathcal{S}_m$, the measurement-induced change in $a_m$ when the conditional state takes the form in Eq.~\eqref{eq:opt_align}. The result is,
\begin{align}
d\mathcal{S}_m &= \sqrt{2 \gamma} \left(\sqrt{a_m a_{m+1}} h_m - a_m \left|\langle J_- \rangle_{{\sf opt},t} \right| \right) dw_{-} \nonumber \\
&\,\,\,\, + 2 \sqrt{\kappa} a_m \left(m - \langle \hat{J}_z \rangle_{{\sf opt},t}\right)dw_z. 
\label{eq:dS}
\end{align}
with
\begin{align}
\left|\langle \hat{J}_- \rangle_{{\sf opt},t}\right| &= \left|\bra{\varphi^{(c)}_{\sf opt}(t)}\hat{J}_- \ket{\varphi^{(c)}_{\sf opt}(t)} \right| =  \sum_{k = -l_T}^{m_T-1} \sqrt{a_k a_{k+1}}h_k\,,\\
\langle \hat{J}_z \rangle_{{\sf opt},t} &= \left|\bra{\varphi^{(c)}_{\sf opt}(t)}\hat{J}_z \ket{\varphi^{(c)}_{\sf opt}(t)} \right| =\sum_{k = -l_T}^{m_T} a_k k \,,
\end{align}
and where $dw_- = \cos(\theta(t)+\pi/2) dw_x + \sin(\theta(t)+\pi/2) dw_y$ is a new independent Wiener increment, obtained via a combination of those from the heterodyne monitoring. 

As a result, we have obtained a system of coupled stochastic differential equations $da_m(t)$
\begin{align}
    da_m &=  \left( -\tilde{u} \sqrt{a_m a_{m+1}} h_m  + \gamma a_{m+1} h_m^2\right)dt  \nonumber \\
    & \,\,\, - \left(-\tilde{u} \sqrt{a_{m-1} a_{m}} h_{m-1}  + \gamma a_{m} h_{m-1}^2\right)dt + d\mathcal{S}_m\,,
    \label{eq:CSME}
\end{align}
that depend solely on the current population levels $\{a_k\}$ and are independent of the control parameter $\theta(t)$. Therefore, the solution to the quantum control problem with the control Hamiltonian from Eq.~\eqref{eq:OPHamiltonian} and the optimal state from Eq.~\eqref{eq:opt_align} corresponds to the solution of a classical stochastic master equation. By numerically simulating this equation, we obtain the values of the corresponding populations at each time $\{a_m(t)\}$ and then, by taking the average over the stochastic realizations and the long time limit,  we can then obtain a {\it conjectured} bound for the steady-state average fidelity
\begin{align}
\overline{\mathcal{F}}_{ss} \stackrel{?}{\leq} \mathscr{B}_\text{DS} \equiv \lim_{t\to\infty} \mathbbm{E}[a_{m_T}(t)]\,.
\end{align}
We will refer to this as the \textit{Dicke-stochastic bound}. 
We remark again that $\mathscr{B}_\text{DS}$ is just a conjecture, based on the assumption that at steady state the optimal conditional state has the form in Eq.~\eqref{eq:opt_align}. We believe that this is indeed the optimal scenario, at least in the long time limit, and we will present numerical evidence in the following sections supporting this conjecture.

Now we have a hierarchy of bounds, each has different levels of generality and computational complexity,
\begin{equation}
    \overline{\mathcal{F}}_{ss} \stackrel{?}{\leq} \mathscr{B_{\text{DS}}} \leq \mathscr{B}_\text{D} \leq \mathscr{B}_\text{QSL}.
    \label{eq:hierarchy}
\end{equation}
$\mathscr{B_{\text{DS}}}$ holds for the control Hamiltonian in Eq.~\eqref{eq:OPHamiltonian}, the monitoring strategy in Eq.~\eqref{eq:SME_Dicke} and a specific value of $\gamma$ and $\kappa$. $\mathscr{B}_\text{D}$ additionally holds as well for the control Hamiltonian in Eq.~\eqref{eq:OPHamiltonian}, but also for any value of $\kappa$ (even if it is conditionally controllable), and for any monitoring strategy that unravels the master equation~\eqref{eq:MEDicke}. Finally, $\mathscr{B}_\text{QSL}$ holds for any Hamiltonian that satisfies Eq.~\eqref{eq:Dicke_var} and for any monitoring strategy (both diffusive and jump-like unravellings). Similarly, the problems get computationally more challenging as we tighten the constraints.
%
%
\section{Optimal Control Strategies}
\label{Sec:opt_cont}
In this section, we consider specific control strategies and compare them to the bounds from the previous section. In all examples, the system evolution is described by Eq.~\eqref{eq:SME_Dicke}.
We will start with the paradigmatic example of the preparation of a qubit state. We will then move to the preparation of Dicke states for $N$ atoms, focusing on the preparation of both the maximally excited state and the maximally entangled Dicke state.
\subsection{The qubit case}
\label{sec:qubit_control}
\begin{figure}[t]
\begin{center}
\includegraphics[angle=0,width=0.98\linewidth]{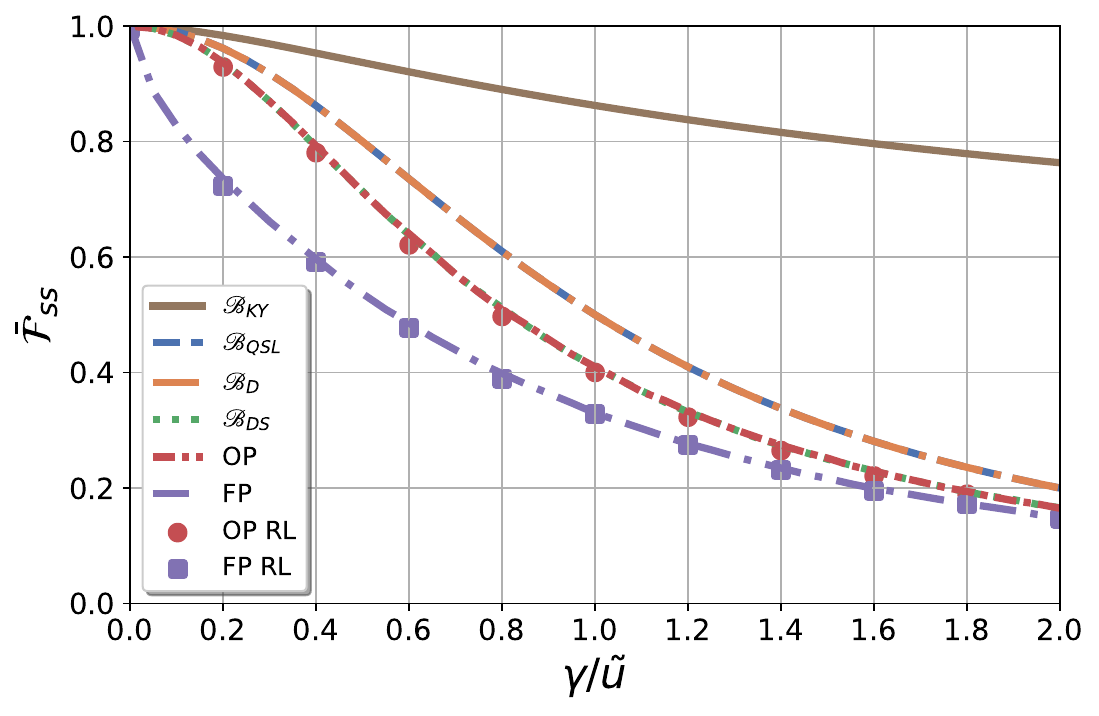} 
\end{center}
\caption{Steady-state fidelity to the excited state of the qubit as a function of $\gamma$ for OP and FP control along with the hierarchy of bounds in Eq.~\eqref{eq:hierarchy}. The results are averaged over 2000 trajectories. 
Parameters: $\kappa/\tilde{u} = 0.4$, $N=1$, $m_T = \frac12$.
}
\label{fig:qubit}
\end{figure}

For simplicity, we start by analyzing the preparation of the excited state of a qubit, thus corresponding to the choice $l_T = m_T = N/2 = 1/2$. Simultaneous monitoring of both the $\sigma_z$ and $\sigma_-$ channels has recently been experimentally demonstrated~\cite{ficheux2018dynamics}. The key property that makes the qubit case simple is that there is only a single relative phase difference $\Delta\phi$ and therefore, the conditional state is always in the optimal form. Therefore, if we consider the optimized-phase control Hamiltonian $\hat{H}_{OP}(t)$ we can always choose $\theta(t) = \Delta\phi + \pi/2$ and we will always saturate Eq.~\eqref{eq:robertson}. Additionally, since there is only a single orthogonal state the QSL upper bound and the Dicke upper bound coincide, i.e. $\mathscr{B}_\text{QSL} = \mathscr{B}_\text{D}$. If we instead consider the fixed-phase control Hamiltonian $\hat{H}_{FP}(t)$, it is not always possible to prepare the optimal conditional state $|\varphi_{\sf opt}^{(c)}(t)\rangle$ at each time $t$. One possible strategy would be to optimize the rate of change of the fidelity locally. For example, if we fix $\theta(t) = \frac{\pi}{2}$ the corresponding coupling strength is
\begin{equation}
    u(t) = - \text{sgn}\left(\Tr\left[\rho^{(c)}\hat{\sigma}_x\right]\right) \Tilde{u}.
    \label{eq:qubitbangbang}
\end{equation}
As we will argue in Sec.~\ref{sec:RLvalidation} this does appear to be the optimal control strategy for $\hat{H}_{FP}(t)$.
We can see our analysis confirmed in Fig.~\ref{fig:qubit}: the optimal control strategy with the $\hat{H}_{OP}(t)$ corresponds exactly to the tighter bound $\mathscr{B}_\text{DS}$. Additionally, $\mathscr{B}_\text{QSL}$ and $\mathscr{B}_\text{D}$ are equal and closely track the behavior of the exact solution. Therefore, the bound indicates the parameter range in which we can reach our desired fidelity without the need for simulation. Finally, we see that phase control is important for obtaining the optimal strategy. The control strategy employed for $\hat{H}_{FP}(t)$ performs significantly worse than the optimized-phase one, especially for small $\gamma/u$ as it is not possible to simultaneously optimize the relative phase and drive in the direction of the target state.

We remark here that, according to the the control strategy described by Eq.~\eqref{eq:qubitbangbang}, the control strength $u(t)$ switches instantaneously between $+\tilde{u}$ and $-\tilde{u}$, depending on the sign of $\text{Tr}[\rho\sigma_x]$ and similar behaviour will be obtained also in the examples we will address in the next sections. This kind of bang-bang control protocol, characterized by abrupt switching between extremal values of the control field, is actually quite common in the quantum control literature, especially in the context of time-optimal or resource-limited strategies. However, as we commented in Sec.~\ref{Sec:measure}, while theoretically efficient, such protocols may challenge the assumptions underlying Markovian dynamics, as the control modulation happens on timescales comparable to or shorter than the bath correlation time. In such instances, one will thus need to smooth the control protocol to ensure consistency with a Lindblad dynamics.

\subsection{Optimal control for Dicke states preparation}
\label{sec:Dicke_control}
As we move to the general $N$ atoms case, the control problem becomes more complicated. This is because we now have $N$ relative phases to deal with and, even with the optimized phase control Hamiltonian $\hat{H}_{OP}(t)$, only a single phase control parameter $\theta(t)$. 
As a consequence, unless the phases are all aligned, such as in the case of the optimal state, the bounds will not be achievable. A common technique to identify a good control strategy is to pick a cost function and locally minimize it, i.e. minimize its time derivative~\cite{stockton2004deterministic}. An example of a cost function we could choose is simply the infidelity, $\mathcal{C} = 1-\mathcal{F}$ or since we have a control Hamiltonian connecting only neighboring Dicke states, we could employ a Euclidean distance measure such as
\begin{equation}
    \label{eq:Euc_cost}
    \mathcal{C} = \sum_m a_m |m-m_T|.
\end{equation}
The rate of change of this cost function due to the control Hamiltonian $\hat{H}_{OP}(t)$ is
\begin{align}
    \frac{d \mathcal{C}}{dt} &= \sum_m [da_m]_{\hat{H}}|m-m_T| \nonumber \\
    &= u \Re\left[e^{i(\theta - \frac{\pi}{2})}\sum_m s(m) \sqrt{a_m a_{m+1}} e^{-i \Delta \phi_m} h_m\right]
    \label{eq:cost_rate}
\end{align}
where $[da_m]_{\hat{H}}$ is derived in Eq.~\eqref{eq:damH} and $s(m) \equiv \text{sign}(m - m_T + 1/2)$. We can locally minimize this cost function by fixing the coupling strength $u(t) = \Tilde{u}$ and setting the phase to
\begin{equation}
    \theta(t) = -\frac{\pi}{2} - \arg\left(\sum_m s(m) \sqrt{a_m a_{m+1}} e^{-i \Delta \phi_m} h_m\right).
    \label{eq:OP_OC}
\end{equation}
For the fixed phase Hamiltonian, $\hat{H}_{FP}(t)$, $\theta = \pi/2$ therefore the minimization occurs when
\begin{equation}
    u(t) = -\text{sign}\left(\Re\left[\sum_m s(m) \sqrt{a_m a_{m+1}} e^{-i \Delta \phi_m} h_m\right]\right) \Tilde{u}.
    \label{eq:FP_OC}
\end{equation} 
A similar cost function, $\mathcal{C} = \sum_m a_m^2 (m - m_T)^2$, was employed in~\cite{stockton2004deterministic} to solve the Dicke state preparation problem in the absence of collective damping ($\gamma=0$). However, we have numerical evidence that the cost function we introduced in Eq.~\eqref{eq:Euc_cost} performs slightly better in practice. In the next sections, we will compare the performances of the optimized OP- and FP-strategies proposed in Eqs.~\eqref{eq:OP_OC} and ~\eqref{eq:FP_OC} with the bounds derived in the manuscript for the two paradigmatic examples of Dicke states: the maximally excited Dicke state $\ket{N/2,N/2}$ and the maximally entangled Dicke state $\ket{N/2,0}$.
\subsection{Preparation of maximally excited Dicke state}
\label{sec:Dicke_excited}
\begin{figure}[t]
\begin{center}
\includegraphics[angle=0,width=0.98\linewidth]{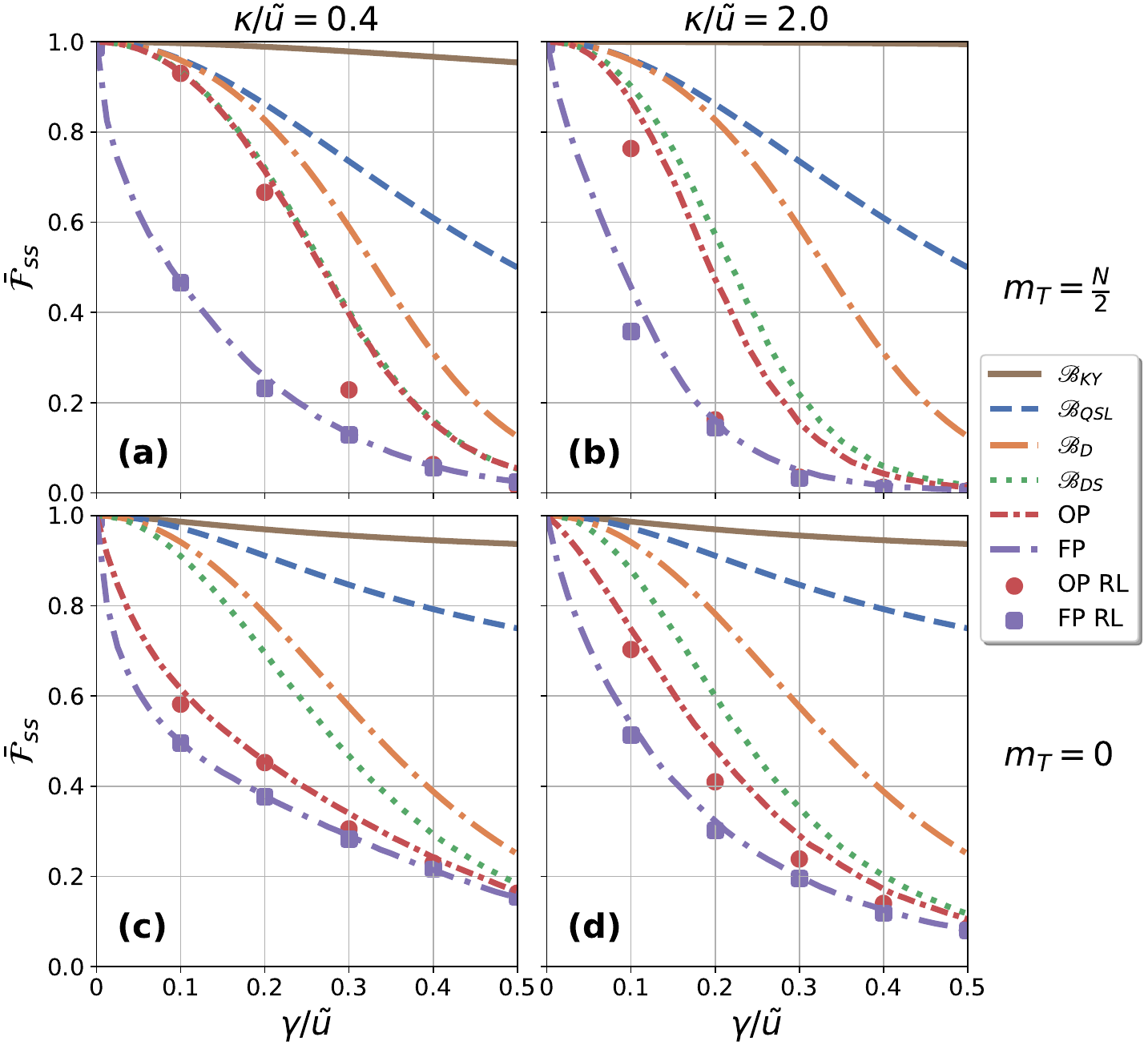} 
\end{center}
\caption{Steady-state fidelity, $\overline{\mathcal{F}}_{ss}$, averaged over 2000 trajectories, as a function of $\gamma/\tilde{u}$ for OP and FP control with $N=4$ along with the hierarchy of bounds in Eq.~\eqref{eq:hierarchy}. 
In \textbf{(a)} and \textbf{(b)} the target state is the maximally excited state $m_T = N/2$ compare to $m_T = 0$ in \textbf{(c)} and \textbf{(d)}. 
The $J_z$ monitoring strength is kept low at $\kappa/\tilde{u} = 0.4$ in \textbf{(a)} and \textbf{(c)} vs $\kappa/\tilde{u} = 2.0$ in \textbf{(b)} and \textbf{(d)}.
The markers represent the steady-state fidelity achieved by the reinforcement learning (RL) agent with access to $\hat{H}_{OP}$ (red, circle) and $\hat{H}_{FP}$ (purple, square) control Hamiltonians.
}
\label{fig:4qubit_compare}
\end{figure}

Let us now focus on the preparation of the maximally excited state $\ket{N/2,N/2}$, that is corresponding to $l_T = m_T = N/2$. In the top panels of Fig~\ref{fig:4qubit_compare} we show the average steady-state fidelities, $\overline{\mathcal{F}}_{ss}$, of the $\sf OP$ and $\sf FP$ control protocols specified in the previous section. In all numerical simulations, we first evolve the system along many trajectories until the unconditional state reaches a steady state. We continue the evolution and sample fidelities, averaging over both time and trajectories to enhance precision by leveraging system ergodicity. These numerical results are compared with the hierarchy of bounds introduced in Sec.~\ref{Sec:Deriv} and plotted as a function of $\gamma/\tilde{u}$ for two values of $\kappa/\tilde{u}$ with $N=4$.
Remarkably, we observe that using the optimized-phase Hamiltonian $\hat{H}_{OP}(t)$ one is able, at least approximately, to simultaneously drive towards the target state and keep the relative phases aligned, saturating $\mathscr{B}_\text{DS}$. However, as expected, this is not possible for the fixed-phase control Hamiltonian $\hat{H}_{FP}(t)$, as shown by the noticeable gap between the two curves. For larger values of $\kappa$ (e.g. $\kappa/\tilde{u}=2.0$) we observe a small gap forming between the OP average fidelity and the Dicke-stochastic bound $\mathscr{B}_\text{DS}$. This can be understood from the fact that monitoring $J_z$ generates spin-squeezing~\cite{thomsenSpinSqueezingQuantum2002}, and thus entanglement which is unwanted in this case since $\ket{N/2,N/2}$ is a separable state. Additionally, in the large $\kappa$ limit the conditional state quickly collapses onto a particular Dicke state resulting in a quantum Zeno effect. 

\begin{figure}[t]
\begin{center}
\includegraphics[angle=0,width=0.99\linewidth]{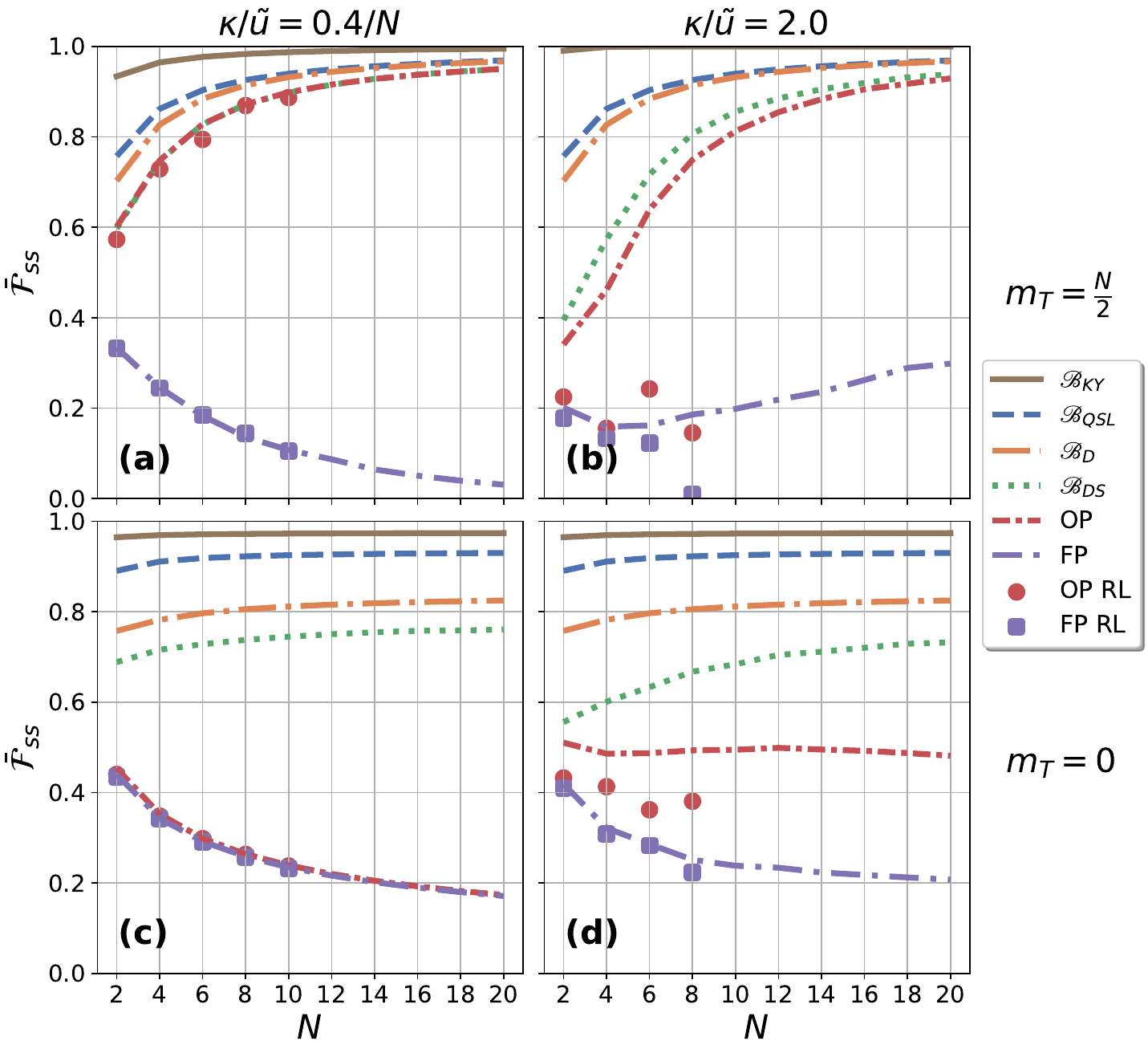} 
\end{center}
\caption{Average steady-state fidelity, $\overline{\mathcal{F}}_{ss}$, as a function of $N$ for OP and FP control with $\frac{N\gamma}{\tilde{u}}=0.8$ along with the hierarchy of bounds in Eq.~\eqref{eq:hierarchy}. 
In \textbf{(a)} and \textbf{(b)} the target state is the maximally excited state $m_T = N/2$ compare to $m_T = 0$ in \textbf{(c)} and \textbf{(d)}. 
The $J_z$ monitoring strength is small and scales inversely with $N$ at $\frac{N\kappa}{\tilde{u}} = 0.4$ in \textbf{(a)} and \textbf{(c)} compared to a large constant value of $\kappa/\tilde{u} = 2.0$ in \textbf{(b)} and \textbf{(d)}. 
The markers represent the steady-state fidelity achieved by the reinforcement learning (RL) agent with access to $\hat{H}_{OP}$ (red, circle) and $\hat{H}_{FP}$ (purple, square) control Hamiltonians. The results are averaged over 1000 trajectories in \textbf{(a)} and \textbf{(c)} and 500 trajectories in \textbf{(b)} and \textbf{(d)}.
} 

\label{fig:DickeN}
\end{figure}

We can now study the behavior of the two strategies as a function of the number of atoms $N$, for a fixed value of $N\gamma/\tilde{u} = 0.8$. The corresponding steady-state fidelities are plotted in Fig.~\ref{fig:DickeN} \textbf{(b)} for a fixed value of $\kappa/\tilde{u}=2.0$, and in Fig.~\ref{fig:DickeN} \textbf{(a)} where we consider the {\it thermodynamical-limit} inverse scaling for the dispersive constant, $N \kappa/\tilde{u} = 0.4$. The control problem revolves around maintaining relative phase alignment and driving toward the desired state. We find it impossible to do both simultaneously with fixed phase control.  In Fig.~\ref{fig:DickeN} \textbf{(b)} the large values $\kappa/\tilde{u}$ hurt the OP control as discussed above. Larger values of $\kappa$ can benefit FP control because it realigns the relative phases.
We also observe that all the bounds capture the scaling behavior of the OP control strategy as $N$ increases with varying degrees of tightness according to the hierarchy in Eq.~$\eqref{eq:hierarchy}$.

If we now focus on the FP control strategy, we do observe in both cases a wide gap between the average steady-state fidelity and all the three bounds we derived; for the larger value of $\kappa$, the average fidelity also exhibits a non-monotonous behaviour as a function of the number of atoms $N$, while for $\kappa$ scaling inversely with $N$, it seems to decrease monotonically towards a finite (but rather small) value.

\subsection{Preparation of the maximally entangled Dicke state}
\label{sec:Dicke_entangled}
We can now analyze the performances of our strategies for the generation of the maximally entangled Dicke state $\ket{N/2,0}$ (corresponding to $l_T=N/2$ and $m_T=0$).

We start by looking at the behavior of the average steady-state fidelities for the two strategies, along with the bounds, as a function of $\gamma/\tilde{u}$ for fixed values of $\kappa$ with $N=4$ (Fig.~\ref{fig:4qubit_compare}). We here observe that, even with the OP control strategy, it is not possible to control the relative phases fully as shown by the gap between the average fidelity and the bound $\mathscr{B}_\text{DS}$. We also find that there is a smaller improvement gained by allowing the additional phase control. Looking at the plots of the average fidelities as a function of the number of atoms $N$ for $\kappa$ fixed (right panel of Fig.~\ref{fig:DickeN}) or for $\kappa$ scaling inversely with $N$ (right panels of  Fig.~\ref{fig:DickeN}), we find non-trivial and interesting behavior. 
A notable feature we find is that when $\kappa$ scales inversely with $N$ there is a negligible amount of entanglement generated in the system, leading to a maximum steady-state fidelity equal to the overlap with the $\ket{+}$ coherent state in the large $N$ limit. This is why the two control curves in Fig.~\ref{fig:DickeN} \textbf{(c)} are identical and so far away from the bounds. Optimizing over $\kappa$ we find that $\kappa/\tilde{u}=2.0$ allows us to generate sufficient entanglement to prepare $\ket{N/2,0}$ while avoiding the Zeno effect seen at large $\kappa$. The result, in Fig.~\ref{fig:DickeN} \textbf{(d)}, is an approximately constant steady state fidelity as a function of $N$. Unfortunately, in the true thermodynamic limit $\kappa/\tilde{u}$ cannot remain constant. We are still unable to approach even the $\mathscr{B}_\text{DS}$ bound for the simple fact that it is impossible to avoid populating the states with angular momentum $m > m_T$ when we only have access to collective control Hamiltonians. The necessity of a large $\kappa/\tilde{u}$ and the fact that it is not easy to stabilizing  $\ket{\psi^\perp}$ in $\ket{N/2,1}$ cause the large gaps to $\mathscr{B}_\text{D}$ and $\mathscr{B}_\text{QSL}$ respectively.

\subsection{Validation of optimal control strategies via reinforcement learning}
\label{sec:RLvalidation}

In Sec.~\ref{sec:Dicke_control} we derived control strategies for both the OP and the FP control Hamiltonians by locally minimizing a cost function $\mathcal{C}$. In this section, we will exploit deep reinforcement learning (DRL) algorithms~\cite{sutton2018reinforcement,mnih2015human} to look for better, non-local strategies, or, as will be the case, validate the optimality of the local strategies. We refer to Appendix~\ref{App:ML} for a brief introduction to DRL, its applications in quantum technologies, and more details on the algorithms we employ.

We start by analyzing the qubit case, where we already know that the OP strategy is indeed optimal, thus allowing us to investigate the effectiveness of our DRL algorithms in finding such an optimal strategy (or at least a different strategy with the same performance). Indeed, as we observe in Fig.~\ref{fig:qubit}, the DRL strategy devised by the neural network attains the same values of steady-state average fidelity of the optimal OP-strategy, and thus also saturates $\mathscr{B}_\text{DS}$. These results clearly show the effectiveness of DRL in finding the optimal strategy. Remarkably, we observe similar results for the FP strategy: the DRL optimized strategy once again attains the same values of fidelity as the locally optimized strategy described in Sec.~\ref{sec:qubit_control}; we remark that this is the first example where we had no previous analytical nor numerical evidence that our strategy was optimal given the constraint of the Hamiltonian. This allows us to conjecture the optimality of the local strategy, implying that no significant advantage can be gained by following a more complicated (non-locally optimized) control strategy.

We can now analyze the scenario with a larger number of atoms ($N>1$) and thus the results shown in Figs.~\ref{fig:4qubit_compare} and~\ref{fig:DickeN} by first considering small values of $\kappa$. As regards the generation of the maximally excited state $\ket{N/2,N/2}$, we can make observations analogous to the ones done for the qubit case: our OP strategy is indeed already optimal (in terms of the steady-state fidelity), as it approaches the bound $\mathscr{B}_{DS}$, and the DRL-optimized strategy attains the same optimal values. This supports the conjecture that $\mathscr{B}_{DS}$ is a valid bound. At the same time, the DRL-optimized results validate our FP control strategy, attaining again the same performance. Similarly, for the maximally entangled Dicke state $\ket{N/2,0}$, when $\kappa$ is small the DRL algorithm achieves very similar results, reinforcing the idea that the steady-state fidelities achieved by the OP and FP strategies are close to the optimal ones.

Although the DRL agent is unable to beat the OP and FP strategies at steady state, our numerical simulations show that the DRL-optimized strategy is in general able to approach the steady-state solution faster than the locally optimized strategies described in the previous sections, particularly when the relative phases of the initial state are unaligned. This is because the reward function values fidelity averaged over the entire trajectory. A different example where the DRL strategy beats OP and FP is seen in Appendix~\ref{App:gamma0} where we analyze the $\gamma \to 0$ limit. In this scenario, the OP and FP strategies fail to converge to the maximally excited state although, convergence can be restored by choosing a different cost function. The DRL strategy sacrifices some fidelity at early times to focus on maintaining phase alignment.  Interestingly, the presence of collective damping helps keep the phases aligned and allows many choices of cost function $\mathcal{C}$ to approach the optimal steady-state fidelity.

If instead, we consider larger values of $\kappa$, we find that the DRL fails in finding optimal strategies, as the corresponding average fidelities are well below our OP and FP strategies. We believe this is because the dephasing term dominates the evolution on a single trajectory level, making it difficult for the DRL agent to extract the reward data from the noise. A different approach that evaluates reward functions beyond the single trajectory level would be more effective in high-noise domains. The totality of the DRL analysis allows us to conjecture the approximate optimality of both the OP and FP strategies for maximizing steady-state fidelity across a wide parameter range.
\section{Conclusions}
\label{sec:conclusions}
In this work, we derived a quantum speed limit (QSL)-inspired upper bound on the fidelity achievable through continuous monitoring and feedback control protocols for quantum systems governed by Markovian master equations. Our results demonstrate that this bound is tighter than previously established bounds in the literature, offering deeper insights into optimal control strategies capable of achieving maximal fidelity. We also extended the bound to incorporate Markovian feedback where the control Hamiltonian strength is statistically unbounded.

Focusing on the preparation of Dicke states in an ensemble of $N$ two-level atoms, we established a hierarchy of increasingly tighter bounds on the steady-state fidelity under collective damping and dispersive coupling. By comparing these bounds to two classes of control strategies—those optimizing the control Hamiltonian phase and those without phase optimization—we highlighted the critical role of phase control, especially in preparing maximally excited states. In the thermodynamic limit ($N \to \infty$), we identified a fundamental limitation: achieving high fidelity requires the collective damping coupling to scale inversely with system size, highlighting a fundamental connection between quantum controllability and thermodynamic-limit descriptions of open quantum dynamics.

 Our theoretical framework also informed the development of practically implementable optimal control strategies via local cost-function optimization. We validated these strategies using deep reinforcement learning (DRL), which consistently converged to similar control strategies and achieved steady-state fidelities comparable to those obtained via our locally optimal approaches. This demonstrates not only the validity but also the practical utility of our bounds in guiding optimal quantum control. 

Our methods are broadly applicable to other quantum state engineering problems. Looking ahead, several promising research directions emerge. Since we have derived a QSL for fidelity—bounding its local rate of change—it is straightforward to extend these bounds to the time-averaged fidelity. While this would introduce dependence on the initial state, it would also lead to a family of differential equations, both stochastic and deterministic, governing the evolution of the time-averaged fidelity. Another intriguing avenue is the extension of our approach to bound quantities beyond fidelity, particularly those that are linear in the quantum state, such as expectation values of observables. Higher-order moments~\cite{bakewell2023} and first-passage\cite{bakewell2024bounds} times are two more areas of interest that could benefit from the ideas presented in this paper.
Finally, in the view of actual experimental implementations of the optimal control protocols we have identified, it will be also useful to study how to smooth the bang-bang control protocols, in order to ensure consistency with a Lindblad dynamics, and to perform a quantitative assessment of how this smoothing affects control performance.
\section{Acknowledgments}
We acknowledge useful discussions with F.~Albarelli, M. Bompais, B.~Huard, A. Smirne, and H. Wiseman.
EO and MGG acknowledge support from MUR and Next Generation EU via the PRIN 2022 Project CONTRABASS (Contract N.2022KB2JJM), NQSTI-Spoke2-BaC project QMORE (contract n. PE00000023-QMORE) and NQSTI-Spoke1-BaC project QSynKrono (contract n. PE00000023-QuSynKrono).
\section*{Appendix}
This Appendix is organized as follows: in Appendix~\ref{App:Drift} we show the details needed to bound the noisy term $\langle \mathscr{D}\rho\rangle_\psi$ as in Eq.~\eqref{eq:noisyterm1}; in Appendix~\ref{app:KYBound} we show how our QSL-bound $\mathscr{B}_\text{QSL}$ is tighter than the one derived by Kobayashi and Yamamoto in~\cite{kobayashi2019control};
in Appendix~\ref{app:mark_fb} we modify the QSL-bound $\mathscr{B}_\text{QSL}$ to include the case of Markovian feedback, and show its usefulness in assessing the performance of a protocol recently proposed in~\cite{campagne2016using};
in Appendix~\ref{app:transferrates} we derive the transfer rates $T_m$ appearing in Eq.~\eqref{eq:transf_rates}; in Appendix~\ref{App:stochastic} we derive the stochastic terms $d\mathcal{S}_m$ in Eq.~\eqref{eq:dS}; 
in Appendix~\ref{App:ML} we give a brief introduction to DRL, to its applications in quantum technologies and we give more details on the algorithms we have exploited to optimize our control strategies;
in Appendix~\ref{App:gamma0} we investigate Dicke state preparation for the special case of $\gamma = 0$.
\appendix
\setcounter{equation}{0}
\renewcommand\theequation{A.\arabic{equation}}

\section{Bounding the noisy term $\langle \mathscr{D}\rho\rangle_\psi$}
\label{App:Drift}
We remind ourselves that every (pure) conditional state can be written in the form $\ket{\varphi^{(c)}} = \sqrt{\mathcal{F}} \ket{\psi} + e^{i \phi^{(c)}} \sqrt{1-\mathcal{F}}\ket{\psi^{\perp(c)}}$ (from now on, to make the equations easier to read, we will omit the $(c)$ superscripts).
Given the fact that $\braket{\varphi}{\psi} = \braket{\psi}{\varphi} = \sqrt{\mathcal{F}}$ we have
\begin{equation}
    \langle \mathcal{D} \rho \rangle_{\psi}
    = \sum_j\left[\lvert\bra{\varphi}\hat{c}_j^\dag\ket{\psi}\rvert^2 - \frac{\sqrt{\mathcal{F}}}{2}\left(\bra{\varphi}\hat{c}_j^\dag\hat{c}_j\ket{\psi} + h.c\right)\right] \,.
\end{equation}
By expanding the first term we obtain
\begin{equation}
\begin{split}
    \lvert\bra{\psi}\hat{c}_j\ket{\varphi}\rvert^2 &= \,\,\mathcal{F} \lvert\!\ev{\hat{c}_j}{\psi}\rvert^2 + (1-\mathcal{F})\lvert \! \mel*{\psi^\perp}{\hat{c}_j^\dag}{\psi}\rvert^2 + \\ \nonumber
    &\quad+ \sqrt{\mathcal{F} - \mathcal{F}^2}\left(e^{-i \phi} \mel*{\psi^\perp}{\hat{c}_j^\dag}{\psi} \expval*{\hat{c}_j}{\psi}+ h.c.\right)\,,
\end{split}
\end{equation}
while the second term yields
\begin{align}
    &\frac{\sqrt{\mathcal{F}}}{2}\left(\bra{\varphi}\hat{c}_j^\dag\hat{c}_j\ket{\psi} + h.c\right) = \\
    &= \mathcal{F}\expval{\hat{c}_j^\dag\hat{c}_j}{\psi} + \frac{\sqrt{\mathcal{F}-\mathcal{F}^2}}{2}\left(e^{-i\phi} \mel*{\psi^\perp}{\hat{c}_j^\dag\hat{c}_j}{\psi} + h.c.\right) \,.
\end{align}
Putting these terms together and then maximizing over $\phi$ we get
\begin{align}
\begin{split}
    \langle \mathcal{D} \rho \rangle_{\psi} &= -\mathcal{F} (\Delta \mathbf{\hat{c}})^2 + (1-\mathcal{F})\sum_j\lvert \! \mel*{\psi^\perp}{\hat{c}_j^\dag}{\psi}\rvert^2\\ 
    & +\frac{\sqrt{\mathcal{F} - \mathcal{F}^2}}{2}\left(e^{-i\phi} \bra*{\psi^\perp}\sum_j \left(2\hat{c}_j^\dag \dyad{\psi} \hat{c}_j - \hat{c}_j^\dag\hat{c}_j\right) \ket*{\psi} + h.c.\right)
\end{split}\nonumber \\
\begin{split}
    &\leq -\mathcal{F} (\Delta \mathbf{\hat{c}})^2 + (1-\mathcal{F})\sum_j\lvert \! \mel*{\psi^\perp}{\hat{c}_j^\dag}{\psi}\rvert^2\\ 
    &\quad +\sqrt{\mathcal{F} - \mathcal{F}^2}\bigg\lvert\bra*{\psi^\perp}\sum_j \left(2\hat{c}_j^\dag \dyad{\psi} \hat{c}_j - \hat{c}_j^\dag\hat{c}_j\right) \ket*{\psi} \bigg\rvert
\end{split}\nonumber \\
&= -\mathcal{F} (\Delta \mathbf{\hat{c}})^2 + (1-\mathcal{F})\mathcal{A}^{(c)} + \mathcal{B}^{(c)}\sqrt{\mathcal{F} - \mathcal{F}^2} \,,
\end{align}
where $(\Delta \mathbf{\hat{c}})^2$, $\mathcal{A}^{(c)}$ and $\mathcal{B}^{(c)}$ are defined respectively in Eqs.~\eqref{eq:Delta_c}, \eqref{eq:A} and \eqref{eq:B}. 

To take the average over trajectories we must maximize $\mathcal{A}^{(c)}$ and $\mathcal{B}^{(c)}$ over all orthogonal states. We can explicitly perform this optimization to obtain a closed form for $\mathcal{A}^*$ and $\mathcal{B}^*$. We introduce the operator $\hat{Q} = \hat{\mathbbm{1}} - \dyad{\psi}$ that satisfies $\hat{Q}\ket*{\psi^\perp} = \ket*{\psi^\perp}$ and $\hat{Q}^2 = \hat{Q}$. We can write $\mathcal{A}$ in the form
\begin{align}
    \mathcal{A}^* &= \max_{\psi_\perp} \sum_j\lvert \! \mel*{\psi^\perp}{\hat{Q} \hat{c}_j^\dag}{\psi}\rvert^2 \nonumber \\ 
    &= \max_{\psi_\perp} \mel*{\psi^\perp}{\hat{Q} \left(\sum_j \hat{c}_j^\dag \dyad{\psi} \hat{c}_j \right) \hat{Q}}{\psi^\perp} \nonumber \\
    & = \lVert \hat{Q} \left(\sum_j \hat{c}_j^\dag \dyad{\psi} \hat{c}_j \right) \hat{Q} \rVert_{\sf op}.
\end{align}
Where $\lVert \hat{O} \rVert_{\sf op}$ represents the operator norm of $\hat{O}$. The following inequality holds as a consequence of Cauchy-Schwarz
\begin{align}
    \lvert \! \mel*{\psi^\perp}{\hat{O}}{\psi} \rvert^2 &= \lvert \! \mel*{\psi^\perp}{\hat{Q}\hat{O}}{\psi}\rvert^2 \nonumber \\
    &\leq \braket*{\psi^\perp}\langle \hat{O}^\dag \hat{Q}\hat{O} \rangle \nonumber \\
    &= \langle \hat{O}^\dag\hat{O} \rangle_{\psi} - \lvert \langle \hat{O} \rangle_\psi \rvert^2 \nonumber \\
    &= (\Delta \hat{O})^2
    \label{eq:perp_inequality}
\end{align}
and is saturable when $\ket*{\psi^\perp} \propto Q \hat{O} \ket{\psi}$. It simply follows that 
\begin{align}
    \mathcal{B}^* &= \max_{\psi_\perp} \bigg\lvert\bra*{\psi^\perp}\sum_j \left(2\hat{c}_j^\dag \dyad{\psi} \hat{c}_j - \hat{c}_j^\dag\hat{c}_j\right) \ket*{\psi} \bigg\rvert \nonumber \\
    &= \Delta \hat{X}_{\mathcal{B}}
\end{align}
where
\begin{equation}
    \hat{X}_{\mathcal{B}} = \sum_j \left(2\hat{c}_j^\dag \dyad{\psi} \hat{c}_j - \hat{c}_j^\dag\hat{c}_j\right)
\end{equation}
This means we now have closed for expressions $\mathcal{A}^*$ and $\mathcal{B}^*$ that depend only on the target state and the jump operators.

\section{Relationship with Kobayashi-Yamamoto bound}
\label{app:KYBound}
\renewcommand\theequation{B.\arabic{equation}}
We can rewrite by using our notation the Kobayashi and Yamamoto bound on the fidelity rate derived in~\cite{kobayashi2019control} as follows
\begin{align}
    \frac{d \overline{\mathcal{F}}}{dt} &\leq - (\Delta \mathbf{\hat{c}})^2 + 2\Delta E \sqrt{1-\overline{\mathcal{F}}}  \nonumber \\
    &+ \sqrt{2 (1-\overline{\mathcal{F}})}\left(\sum_j \langle\hat{c}_j\hat{c}_j^\dag\rangle_{\psi} + \sqrt{\langle(\hat{c}_j^\dag\hat{c}_j)^2\rangle_{\psi}}\right)
    \label{eq:KYbound}
\end{align}
From now on we will suppress $\psi$ in the expectations for readability. To show that our bound is tighter we must show that the RHS of the inequality above is larger than the RHS of Eq.~\eqref{eq:fid_rate_sol}. Due to the tighter QSL-inspired inequality we apply to our Hamiltonian term, it is strictly smaller than the corresponding term in Eq.~~\eqref{eq:KYbound}, therefore we can cancel these terms. Next, we add $(\Delta \mathbf{\hat{c}})^2$ to the each side. The RHS of Eq.~\eqref{eq:fid_rate_sol} is now in the form
\begin{align}
    &\quad(1-\overline{\mathcal{F}})((\Delta \mathbf{\hat{c}})^2 + \mathcal{A}^*) + \mathcal{B}^* \sqrt{\overline{\mathcal{F}} (1-\overline{\mathcal{F}})} \nonumber \\
    &\leq \sqrt{(1-\overline{\mathcal{F}})} \left((\Delta \mathbf{\hat{c}})^2 + \mathcal{A}^* + \mathcal{B}^*\right).
    \label{eq:fid_rate_sol_bound}
\end{align}
The RHS sides of both Eqs.~\eqref{eq:KYbound} and \eqref{eq:fid_rate_sol_bound} have the same $\sqrt{1-\overline{\mathcal{F}}}$ prefactor so our new inequality we want to prove is
\begin{equation}
    \left((\Delta \mathbf{\hat{c}})^2 + \mathcal{A}^* + \mathcal{B}^*\right) \leq \sqrt{2} \left( \sum_j \langle\hat{c}_j\hat{c}_j^\dag\rangle + \sqrt{\langle(\hat{c}_j^\dag\hat{c}_j)^2\rangle} \right)
    \label{eq:first_KY_ineq}
\end{equation}
Applying Eq.~\eqref{eq:perp_inequality} to each term in $\mathcal{A}$ we find that
\begin{equation}
\mathcal{A}^* \leq \sum_j \langle \hat{c}_j\hat{c}_j^\dag \rangle - \lvert \langle \hat{c}_j \rangle \rvert^2 = \sum_j (\Delta\hat{c}_j^\dag)^2.
\end{equation}
To tackle $\mathcal{B}^*$ we will split it up into two terms and apply Eq.~\eqref{eq:perp_inequality} to each
\begin{align}
    \mathcal{B}^* &= \bigg\lvert\bra*{\psi^\perp}\sum_j \left(2\hat{c}_j^\dag \dyad{\psi} \hat{c}_j - \hat{c}_j^\dag\hat{c}_j\right) \ket*{\psi} \bigg\rvert \nonumber \\
    &\leq \sum_j \bigg\lvert\bra*{\psi^\perp} 2\hat{c}_j^\dag \dyad{\psi} \hat{c}_j\ket*{\psi}\bigg\rvert + \bigg\lvert\bra*{\psi^\perp} \hat{c}_j^\dag\hat{c}_j \ket*{\psi} \bigg\rvert \nonumber \\
    &\leq \sum_j 2 (\Delta\hat{c_j}^\dag) \lvert \langle\hat{c}_j\rangle\rvert+ \sqrt{\langle (\hat{c}_j^\dag\hat{c}_j)^2 \rangle - \langle \hat{c}_j^\dag\hat{c}_j \rangle^2} \nonumber \\
    & \leq \sum_j 2 (\Delta\hat{c_j}^\dag) \lvert \langle\hat{c}_j\rangle\rvert + \sqrt{2}\sqrt{\langle (\hat{c}_j^\dag\hat{c}_j)^2 \rangle} - \langle \hat{c}_j^\dag\hat{c}_j \rangle.
\end{align}
In the final line, we used the inequality $\sqrt{x^2-y^2} \leq \sqrt{2}x-y$. Each term is now a sum over $j$ so we just have to prove the following inequality for a single $j$:
\begin{align}
(\Delta\hat{c}_j)^2 + (\Delta\hat{c}_j^\dag)^2 + 2 (\Delta\hat{c_j}^\dag) \lvert \langle\hat{c}_j\rangle\rvert + \sqrt{2}\sqrt{\langle (\hat{c}_j^\dag\hat{c}_j)^2 \rangle} - \langle \hat{c}_j^\dag\hat{c}_j \rangle \nonumber \\
\leq \sqrt{2}\left( \langle\hat{c}_j\hat{c}_j^\dag\rangle + \sqrt{\langle(\hat{c}_j^\dag\hat{c}_j)^2\rangle}\right)
\label{eq:new_thesis}
\end{align}

We can cancel $\sqrt{2}\sqrt{\langle (\hat{c}_j^\dag\hat{c}_j)^2 \rangle}$ on each side of Eq.~\eqref{eq:new_thesis}. We also see that $(\Delta \hat{c}_j)^2 - \langle \hat{c}_j^\dag\hat{c}_j\rangle = -\lvert \langle \hat{c}_j \rangle \rvert^2$. Combining everything together we can restate the inequality we need to prove in the form
\begin{align}
     (\Delta\hat{c_j}^\dag)^2 +  2 (\Delta\hat{c_j}^\dag)  \lvert \langle\hat{c}_j\rangle\rvert - \lvert \langle \hat{c}_j \rangle \rvert^2 &\leq  \sqrt{2} \langle\hat{c}_j\hat{c}_j^\dag\rangle \,, \nonumber 
\end{align}
or equivalently
\begin{align}
    (\Delta\hat{c_j}^\dag)^2 +  2(\Delta\hat{c_j}^\dag)  \lvert \langle\hat{c}_j\rangle\rvert - \lvert \langle \hat{c}_j \rangle \rvert^2 &\leq
      \sqrt{2} \left((\Delta\hat{c_j}^\dag)^2 + \lvert \langle \hat{c}_j \rangle \rvert^2\right) \,.
\end{align}
Moving all terms to the right-hand side, we observe that a perfect square can be reconstructed 
\begin{align}
    0 &\leq \left(\!\sqrt{2}-1\right)\left(\Delta\hat{c_j}^\dag\right)^2 - 2\left(\Delta\hat{c_j}^\dag\right)\lvert \langle\hat{c}_j\rangle\rvert  + \left(\!\sqrt{2}+1\right)\lvert \langle \hat{c}_j \rangle \rvert^2  \nonumber \\
    &= \left(\sqrt{\!\sqrt{2}-1}\left(\Delta \hat{c}_j^\dag\right) - \sqrt{\!\sqrt{2}+1}\lvert \langle \hat{c}_j \rangle \rvert\right)^2 \,,
\end{align}
and thus finally proving that our bound is always tighter than the KY-bound. The main technique that results in the improved tightness of our bound is performing the explicit maximization over $\ket{\psi^\perp}$ instead of relying on norm-based inequalities.

\section{Markovian feedback bound}
\label{app:mark_fb}
\renewcommand\theequation{C.\arabic{equation}}
We will derive a bound on the steady-state fidelity when Markovian feedback is employed as in Eq.~\eqref{eq:Mark_Ham} when the current, $I(t)$ is measured over a finite time $\Delta t$. To do this, we start by bounding the Hamiltonian contribution in an identical manner to Eq.~\eqref{eq:robertson}
\begin{align}
    \label{eq:robertson_mark}
     \langle\mathcal{U}\rho\rangle_{\psi} &=  -i\ev{[\hat{H}^{(c)}, \rho^{(c)}]}{\psi}  \nonumber \\
    &\leq \left| \!\ev{[\hat{H}^{(c)}, \rho^{(c)}]}{\psi} \right| \nonumber \\
    &\leq 2 |I(t)|\Delta \hat{F} \sqrt{\mathcal{F} - \mathcal{F}^2}
\end{align}
The problem remains that $|I(t)|$ is correlated with $\mathcal{F}$. We will tackle this by splitting the current into its two components and bounding each individually,
\begin{equation}
    \frac{1}{\Delta t}\int_{t-\Delta t}^{t}dt' \sqrt{\eta} \tr[\rho^{(c)}(\hat{c} +\hat{c}^\dag)] \leq \lVert\hat{c} +\hat{c}^\dag\rVert_{\sf op}.
\end{equation}
We remark that if $\lVert\hat{c} +\hat{c}^\dag\rVert_{\sf op}$ is infinite, a more sophisticated approach will be necessary that accounts for the structure of the feedback Hamiltonian. This is now uncorrelated with $\mathcal{F}$, so we can average over $\sqrt{\mathcal{F} - \mathcal{F}^2}$ using Jensen's inequality as in Eq.~\eqref{eq:upperbound_unitary}. The second component takes the form
\begin{equation}
    \frac{1}{\Delta t}\int_{t-\Delta t}^{t} dw = \frac{1}{\Delta t}\mathcal{N}(0,\Delta t) = \frac{1}{\sqrt{\Delta t}} \, \mathcal{N}(0,1) \equiv \frac{\Delta w}{\Delta t}.
\end{equation}
$\Delta w$ and $\rho^{(c)}$ are still correlated because the conditional state at time $t$ depends on the measurement record. However, we can get around this by first applying Cauchy-Schwarz followed by Jensen's inequality
\begin{align}
    &\quad \mathbbm{E}[|\Delta w|\sqrt{\mathcal{F} - \mathcal{F}^2}] \nonumber \\
    &\leq \sqrt{\mathbbm{E}[|\Delta w|^2]\mathbbm{E}[\mathcal{F} - \mathcal{F}^2]} \nonumber \\
    &\leq \sqrt{\Delta t} \textstyle\sqrt{\,\overline{\mathcal{F}} - \smash{\overline{\mathcal{F}}}^2}
\end{align}
Combining everything, we obtain the bound
\begin{equation}
    \label{eq:robertson_mark_avg}
     \overline{\langle\mathcal{U}\rho\rangle}_{\psi} \leq  2 \left(\lVert\hat{c} +\hat{c}^\dag\rVert_{\sf op}  + \frac{1}{\sqrt{\Delta t}}\right) \Delta \hat{F} \textstyle\sqrt{\overline{\mathcal{F}} - \smash{\overline{\mathcal{F}}}^2}
\end{equation}
The rest of the analysis in Sec. \ref{Sec:Deriv} also applies to the Markovian case, therefore we can derive an analogous bound to Eq.~\eqref{eq:fid_rate_sol} with
\begin{equation}
    \Delta E \to \left(\lVert\hat{c} +\hat{c}^\dag\rVert_{\sf op}  + \frac{1}{\sqrt{\Delta t}}\right) \Delta \hat{F}\,.
\end{equation}
However, we are interested in state stabilization so we will focus on the regime where $\Delta t$ is very small compared to other timescales in the dynamics. In this limit, we can approximate the steady-state fidelity bound as
\begin{equation}
    \mathcal{F}_{ss} \lessapprox 1 - \left(\frac{(\Delta \mathbf{\hat{c}})^2}{2 \Delta \hat{F}}\right)^2\Delta t + \mathcal{O}(\Delta t^2).
\end{equation}
Therefore, achieving an average stabilization $1 - \mathcal{F}_{ss} \leq \varepsilon$ requires a vanishingly small measurement window
\begin{equation}
    \Delta t \leq \varepsilon \left(\frac{2 \Delta \hat{F}}{(\Delta \mathbf{\hat{c}})^2}\right)^2
\end{equation}
and a diverging average Hamiltonian strength
\begin{align}
    \mathbbm{E}[\Delta \hat{H}^{(c)}] &= 2\mathbbm{E}[|I(t)|] \Delta \hat{F} \nonumber \\
    &\geq \frac{2}{\Delta t}\mathbbm{E}[|\Delta w|] \Delta \hat{F} \nonumber \\
    &= 2 \sqrt{\frac{2}{\pi \Delta t}}\Delta \hat{F} \nonumber \\
    &\geq \sqrt{\frac{2}{\pi \varepsilon}}(\Delta \mathbf{\hat{c}})^2
\end{align}

\begin{figure}[t]
\begin{center}
\includegraphics[angle=0,width=0.98\linewidth]{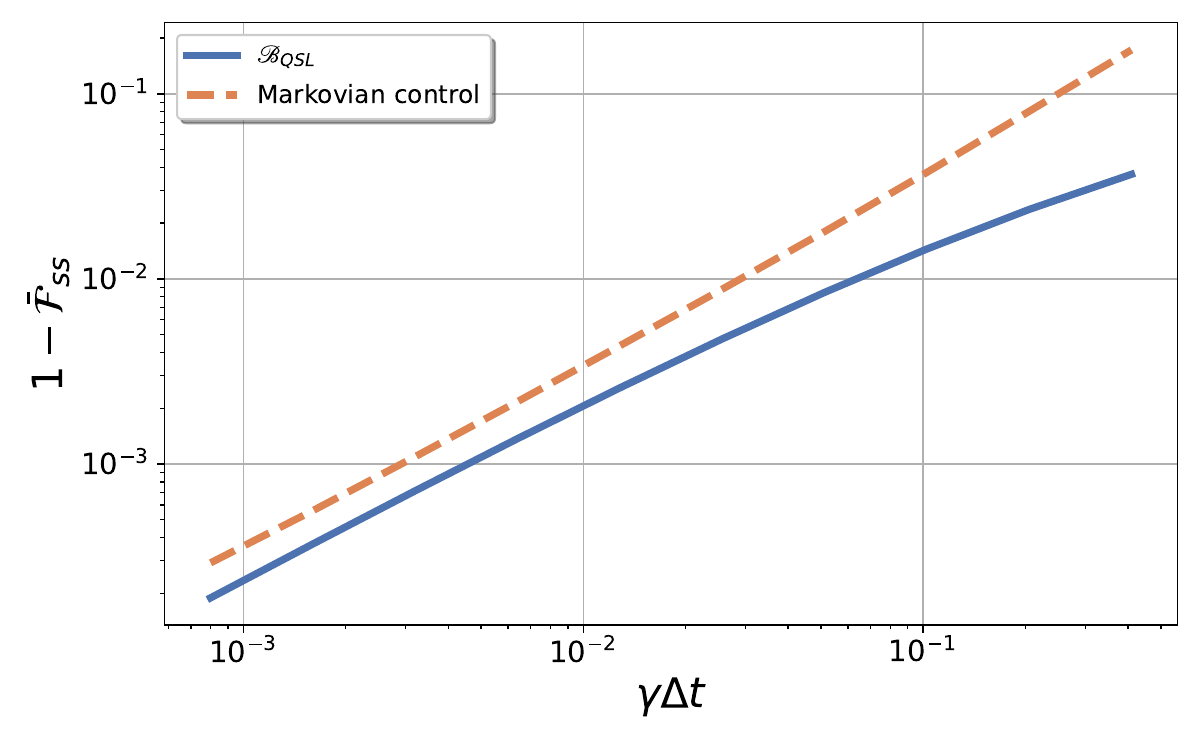} 
\end{center}
\caption{The steady state infidelity (orange, dashed), averaged over 2000 trajectories for stabilization of a qubit in the excited state in the presence of damping. The state is stabilized using Markovian feedback control where the measurement current is averaged over a time interval $\Delta t$. This control strategy is compared to the bound derived in Eq.~\eqref{eq:mark_qubit_bound}, (blue, solid).}
\label{fig:markovian_qubit}
\end{figure}

We illustrate the scaling behavior of this bound by attempting to stabilize the excited state of a qubit in the presence of damping using the method proposed in Ref.~\cite{campagne2016using}. The Stochastic master equation takes the following form
\begin{equation}
d\rho^{(c)} = - i[\hat{H}, \rho^{(c)}]   + \gamma\mathcal{D}[\hat{\sigma}_-]\rho^{(c)} + \sqrt{\eta \gamma} \mathcal{H}[\hat{\sigma}_-]\rho^{(c)} dw
\end{equation}
The authors prove that when $\eta = 1$ choosing $\hat{F} = -\sqrt{\gamma}\sigma_y$ results in the unconditional dynamics of the system stabilizing in the excited state. The relevant terms for our bound are
\begin{align}
    \Delta \hat{F} = 2\sqrt{\gamma} \Delta \sigma_y &= 2\sqrt{\gamma}\\
    \lVert\sqrt{\gamma}(\hat{\sigma}_- +\hat{\sigma}_+)\rVert_{\sf op} &= \sqrt{\gamma} \\
    (\Delta \mathbf{\hat{c}})^2 &= \gamma
\end{align}
Since $\mathcal{A}^* = \mathcal{B}^* = 0$ we can use Eq.~\eqref{eq:BQ_simple} to find
\begin{equation}
    1-\overline{\mathcal{F}}_{ss} \geq \frac{\gamma \Delta t}{4+ 8 \sqrt{\gamma \Delta t} + 5 \gamma \Delta t} 
    \label{eq:mark_qubit_bound}
\end{equation}
In Fig~\ref{fig:markovian_qubit} we compare the bound with a numerical simulation of the Markovian feedback specified above with a finite measurement window. Remarkably, despite the additional inequalities applied in the derivation of this bound, it accurately captures both the magnitude and scaling behavior of the protocol, particularly in the small $\Delta t$ limit.

\section{Derivation of transfer rates $T_m$}
\label{app:transferrates}
\renewcommand\theequation{D.\arabic{equation}}
Assuming a pure conditional state $\rho^{(c)}=\ket{\varphi^{(c)}}\bra{\varphi^{(c)}}$ as in Eq.~\eqref{eq:pure_decomp},  the contribution given by the OP-control Hamiltonian to $d a_m$ can be written as 
\begin{align}
   \left[ d a_m \right]_{\sf \hat{H}} &= -i \bra{l,m}[\hat{H}_{OP}, \rho]\ket{l,m} \nonumber \\
    &= -i e^{-i\phi_m} \sqrt{a_m} \bra{l,m}\hat{H}_{OP} \ket{\varphi^{(c)}} + h.c.
\end{align}
By rewriting the Hamiltonian $\hat{H}_{OP}$ as 
\begin{equation}
    \hat{H}_{OP} = \frac{{u}}{2}\left(e^{-i\theta}\hat{J}_+ + e^{i\theta}\hat{J}_- \right)
\end{equation}
we can easily prove that
\begin{equation}
    \bra{l,m}\hat{H}_{OP} \ket{\varphi^{(c)}} = \frac{{u}}{2}\left(e^{i(\phi_{m-1} - \theta)} h_{m-1}\sqrt{a_{m-1}} + e^{i(\phi_{m+1} + \theta)} h_{m}\sqrt{a_{m+1}}\right)
\end{equation}
and finally,
\begin{align}
    \left[ da_m  \right]_{\sf \hat{H}} &= \frac{{u}}{2} \left(e^{i \alpha_m} \sqrt{a_m a_{m+1}} h_m - e^{i \alpha_{m-1}} \sqrt{a_{m-1} a_m} h_{m-1}\right) + h.c \nonumber \\
    &= u \left(\sqrt{a_m a_{m+1}} h_m \cos(\alpha_m) - \sqrt{a_{m-1} a_m} h_{m-1} \cos(\alpha_{m-1})\right)\,,
    \label{eq:damH}
\end{align}
where $\alpha_m = \theta(t) - \Delta \phi_m - \pi/2$ and $\Delta \phi_m \equiv \phi_m - \phi_{m+1}$.

As the operator $\hat{J}_z$ has no effect on $a_m$ on average, the contribution from the noisy term can simply be obtained by calculating 
\begin{align}
    \left[ da_m  \right]_{\mathcal{D}}= \expval{\mathcal{D}[\gamma \hat{J}_-]\rho}{l,m} dt \,.
\end{align}
The first term in the Lindblad superoperator gives
\begin{align}
    \gamma \expval{\hat{J}_-\rho \hat{J}_+}{l,m} &= \gamma \lvert \bra{l,m} \hat{J}_- \ket{\varphi} \rvert^2 \nonumber \\
    &= \gamma a_{m+1} h_m^2
\end{align}
while for the second term, we can use the fact that $\hat{J}_+\hat{J}_- = \hat{\mathbf{J}}^2 - \hat{J}_z^2 + \hat{J}_z$ and thus
\begin{align}
    &\frac{\gamma}{2}\left( \expval{\hat{J}_+ \hat{J}_- \rho}{l,m} + \expval{\rho \hat{J}_+ \hat{J}_-}{l,m}\right) \nonumber \\
    \quad &= \frac{\gamma}{2} \bra{l,m} {\hat{\mathbf{J}}^2 - \hat{J}_z^2 + \hat{J}_z \rho} \ket{\varphi} \braket{\varphi}{l,m} + h.c \nonumber \\ 
    \quad &= \gamma a_m h_{m-1}^2 \,.
\end{align}
Combining the two results we get
\begin{equation}
    \left[ da_m \right]_{\mathcal{D}}= \gamma a_{m+1} h_m^2 dt - \gamma a_m h_{m-1}^2 dt .
\end{equation}
As a consequence, by adding the Hamiltonian term and the noisy term we obtain the differential equation
\begin{align}
    da_m &=   \left[ da_m \right]_{\hat{H}} +  \left[ da_m \right]_{\mathcal{D}} + d\mathcal{S}_m \nonumber \\
    &=\left( u \sqrt{a_m a_{m+1}} \cos(\alpha_m) h_m  + \gamma a_{m+1} h_m^2\right)dt  \nonumber \\
    & \,\,\, - \left(u \sqrt{a_{m-1} a_{m}} \cos(\alpha_{m-1}) h_{m-1}  + \gamma a_{m} h_{m-1}^2\right)dt \nonumber \\ & \,\,\, + d\mathcal{S}_m\,,
\end{align}
where $d\mathcal{S}_m$ represents the terms coming from the stochastic part of the evolution, and where we have highlighted the fact that the deterministic terms allow transfer only between neighboring Dicke states; we can thus identify the transfer rate into the level $m$ from the level $m+1$ as
\begin{align}
T_m = u \sqrt{a_m a_{m+1}} \cos(\alpha_m) h_m +  \gamma a_{m+1} h_m^2 \,.
\end{align}
\section{Derivation of stochastic terms $d\mathcal{S}_m$}
\renewcommand\theequation{E.\arabic{equation}}
\label{App:stochastic}
We derive here the stochastic terms $d\mathcal{S}_m$ corresponding to the monitoring strategy described by the stochastic master equation \eqref{eq:SME_Dicke}. We start by considering the contribution given by the first term due to the heterodyne detection, i.e.
\begin{align}
    \left[ d a_m \right]_{\sf het,x} &= \ev{\mathcal{H}[\sqrt{\gamma/2} \hat{J}_-]\rho^{(c)}}{l,m} dw_x \nonumber \\
    &= \sqrt{2 \gamma} \left(\sqrt{a_m a_{m+1}}\cos(\Delta \phi_m) h_m - a_m\Re[\langle \hat{J}_- \rangle_\varphi]\right) dw_x.
\end{align}
where $\Delta \phi_m = \phi_m - \phi_{m+1}$, and $\langle J_- \rangle_\varphi=\bra{\phi^{(c)}}\hat{J}_-\ket{\phi^{(c)}}$. The second term corresponding to heterodyne measurement is
\begin{align}
    \left[ d a_m \right]_{\sf het,y}&= \ev{\mathcal{H}[i\sqrt{\gamma/2} \hat{J}_-]\rho^{(c)}}{l,m} dw_y \nonumber \\
    &= \sqrt{2 \gamma} \left(\sqrt{a_m a_{m+1}}\sin(\Delta \phi_m) h_m + a_m\Im[\langle \hat{J}_- \rangle_\varphi]\right) dw_y.
\end{align}
Following the analysis in Sec.~\ref{Sec:BDS} we assume the conditional state is always in the form found in Eq.~\eqref{eq:opt_align}, therefore, we have
\begin{equation}
    \langle \hat{J}_- \rangle_\varphi = e^{- i (\theta+\frac{\pi}{2})} \sum_{k = -l}^{m_T-1} \sqrt{a_k a_{k+1}}h_k \equiv \langle \hat{J}_- \rangle_{{\sf opt},t} \,.
\end{equation}
Since in this state $\Delta \phi_m = \theta(t) + \frac{\pi}{2}$, we can combine the two contributions above and exploit the properties of the two independent Wiener increments, to get a term that is independent of $\theta(t)$
\begin{align}
    \left[d a_m\right]_{\sf het} &= \left[ d a_m \right]_{\sf het,x} + \left[ d a_m \right]_{\sf het,y} \nonumber \\
    &=\sqrt{2 \gamma} \left(\sqrt{a_m a_{m+1}} h_m - a_m\left|\langle \hat{J}_- \rangle_{{\sf opt},t}\right| \right) dw_{-}\,,
\end{align}
where $dw_- = \cos(\theta(t)+\pi/2) dw_x + \sin(\theta(t)+\pi/2) dw_y$ is a new independent Wiener increment.

Similarly, we can derive the contribution due to the continuous homodyne measurement of the $\hat{J}_z$ channel, which yields
\begin{align}
   \left[d a_m \right]_{\sf hom,z} &= \ev{\mathcal{H}[\sqrt{\kappa} \hat{J}_z]\rho^{(c)}}{l,m} dw_z \nonumber \\
    &= 2 \sqrt{\kappa} a_m (m - \langle \hat{J}_z \rangle_{{\sf opt},t})dw_z.
\end{align}
where
\begin{equation}
    \langle J_z \rangle_{{\sf opt},t} = \bra{\varphi^{(c)}_{\sf opt}(t)}\hat{J}_z \ket{\varphi^{(c)}_{\sf opt}(t)} = \sum_{k=-l_T}^{m_T} a_k k.
\end{equation}
We can thus finally write the stochastic contribution to the evolution of the Dicke populations $a_m$ as 
\begin{align}
d\mathcal{S}_m &= \left[d a_m \right]_{\sf het} + \left[d a_m \right]_{\sf hom,z} \nonumber \\
&= \sqrt{2 \gamma} \left(\sqrt{a_m a_{m+1}} h_m - a_m \left|\langle J_- \rangle_{{\sf opt},t} \right| \right) dw_{-} \nonumber \\
&\,\,\,\, + 2 \sqrt{\kappa} a_m \left(m - \langle \hat{J}_z \rangle_{{\sf opt},t}\right)dw_z. 
\end{align}
This means that we are left with a classical stochastic master equation in terms of the Dicke state populations, $\{a_m\}$, that can be simulated at finite time or solved via the Fokker-Planck equation for a steady state solution.

\section{Deep reinforcement learning for quantum state engineering}
\renewcommand\theequation{F.\arabic{equation}}
\label{App:ML}

\begin{figure}[h]
\begin{center}
\includegraphics[angle=0,width=0.98\linewidth]{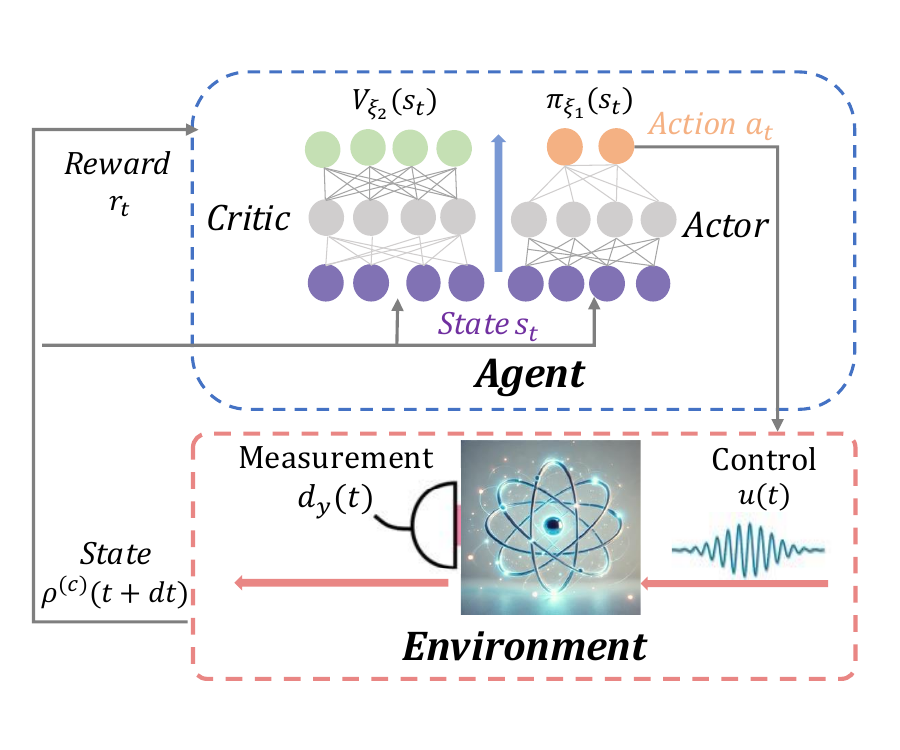} 
\end{center}
\caption{RL pipeline for optimal control of an open quantum system. The bottom box represents the state evolution of the quantum system, and the top box acts as an \textit{agent} to suggest good control based on current knowledge of the quantum state. At timestep $t$, the density matrix of a quantum state $\rho(t)$ (in this case the conditional state $\ket{\varphi^{(c)}}$) is represented as \textit{state}. Then, the \textit{actor} $\pi$ outputs a suggested \textit{action} as $a_t=\pi(s_t)$, and the \textit{critic} evaluates $V(s_t)$, which is utilized to obtain the advantage function $\hat{A}_t=G_t-V(s_t)$. The control field $u(t)$ originating from $a_t$ is performed to evolve the state and the measurement process further evolves the quantum state towards the conditional state $\rho^{(c)}(t+dt)$. \textit{Reward} signals are obtained based on fidelity evaluation of current states, which is utilized to optimize both the parameters of \textit{actor} and \textit{critic}.}
\label{fig:RL}
\end{figure}

Machine learning has attracted increasing attention owing to its powerful computing capability and has been applied in various quantum tasks in recent years. In particular, reinforcement learning (RL)~\cite{sutton2018reinforcement} offers a considerable advantage in controlling systems without prior knowledge about the system model. Deep reinforcement learning (DRL), which combines RL and deep learning, has been recognized as a universal data-driven tool for complex systems~\cite{mnih2015human} and has achieved efficient control of different quantum systems~\cite{zhang2019does,mackeprang2019reinforcement,an2019deep,niu2019universal,fosel2018reinforcement,xu2019generalizable,chen2019extreme,haug2020classifying,ding2021breaking,Borah2021,Brown_2021,Fallani2022,porotti2022deep,sivak2023real,stevetutorial2025,ma2022curriculum,zhou2025auxiliary} (see in particular Refs.~\cite{fosel2018reinforcement,stevetutorial2025,ma2025} for an overview on this topic). Proximal Policy Optimization (PPO)~\cite{schulman2017proximal} has emerged as one of the most powerful DRL variants and has widely applied to different quantum control tasks~\cite{chen2019extreme,haug2020classifying,Brown_2021,ding2021breaking,Borah2021,Fallani2022,porotti2022deep,sivak2023real}. 

Finding optimal feedback strategies for continuously monitored quantum systems is the perfect playground for RL. The stochastic nature of the problem, combined with the presence of feedback mechanisms, results in a double-exponential growth of the space of possible strategies as the number of time steps increases. This complexity places the task beyond the reach of standard optimal control and supervised learning techniques. However, it aligns naturally with the RL paradigm, where an agent explores the problem space through random experiments while simultaneously learning an optimal policy.

RL learns as follows: an \textit{agent} observes its \textit{environment}, takes an \textit{action} based on that observation (\textit{state}), and receives a scalar value \textit{reward} from the \textit{environment}. The mapping from \textit{state} to \textit{action} is defined as the control strategy, known as a \textit{policy}~\cite{sutton2018reinforcement}. At each time step, the \textit{state} is denoted as $s_t$, the \textit{action} as $a_t$, and the \textit{reward} as $r_t$. The goal of RL is to maximize the expected accumulated reward across different trajectories, defined as $G_T = \mathbb{E} \left[ \sum_{t=0}^{T} \gamma^t r_t \right]$, where $T$ is a final time step, $\gamma \in [0,1] $ is the discount factor that controls the trade-off between immediate and future rewards~\cite{sutton2018reinforcement}. Policy-based RL methods directly define the policy as $a_t=\pi(s_t)$ (known as an \textit{actor}). By contrast, value-based RL methods optimize value functions e.g., $Q(s_t,a_t)$ or $V(s_t)$ (known as a \textit{critic}), and the optimal policy can be obtained from the value function, e.g., $a_t^* = \arg\max_{a} Q(s_t, a)$. Furthermore, the actor-critic approach combines value-based and policy-based methods by learning both the value and the policy function, contributing to more stable and efficient learning. 

In our case, we want to control the quantum system based on the results obtained via continuous monitoring. The goal is to prepare some desired quantum state and stabilize it against noise, e.g., from measurement backaction itself and from external decay or decoherence. Hence, the \textit{environment} is modeled using a simulation of the physical time evolution following the SME in Section~\ref{Sec:Deriv}. We use the measurement record to evolve the quantum state according to the SME as was done in~\cite{Fallani2022,porotti2022deep}. We then feed the output of the SME, the density matrix $\rho^{(c)}(t)$ or in this case the conditional state $\ket{\varphi^{(c)}}$ as the \textit{state} of RL. The state was parameterized into a vector containing the real and imaginary part of each Dicke state amplitude. Meanwhile, the \textit{action} is represented by  $a_t = \left[u(t),\theta(t)\right]$, that is by both the control strength $u(t)$ and the phase $\theta(t)$ for the OP-control strategy, and simply as the control strength $a(t)=[u(t)]$ for FP-control. The procedure of finding the optimal control fields that drive the quantum system towards a target state using PPO is summarized in Fig.~\ref{fig:RL}. The \textit{environment} (the bottom box) is simulated using SME, and the agent (the top box) contains an \textit{actor} that suggests an \textit{action} based on the \textit{state} and a \textit{critic} that predicts the \textit{value} which is utilized to obtain the advantage function. In particular, at each time step $t$ with quantum state as $\rho^{(c)}(t)$ (with $\rho^{(c)}(0)=\rho(0)=|\psi_0\rangle \langle \psi_0|$), the \textit{actor} suggests an action $a(t)=\pi_{\xi_1}(s_t)$ with $\xi_1$ denoting the parameters in the \textit{actor} network. Meanwhile, the \textit{critic} evaluates the $V_{\xi_2}(s_t)$ with $\xi_2$ denoting the parameters in the \textit{critic} network. $V(s_t)$ is useful for computing the advantage function $\hat{A}_t=G_t-V(s_t)$ ~\cite{schulman2017proximal}. To avoid big policy updates, PPO employs a constrained loss function to update the parameters, defined as  
\begin{equation}
L^{PPO}(\xi_1) = \mathbb{E}_t \left[ \min \left( p_t(\xi_1)  \hat{A}_t, \text{clip} \left( p_t(\xi_1), 1 - \epsilon, 1 + \epsilon \right) \hat{A}_t \right) \right],
\end{equation}
where $p_t(\xi_1)=\frac{\pi_{\xi_1}(a_t|s_t)}{\pi_{\xi_1^{\prime}}(a_t|s_t)}$, $\xi_1^{\prime}$ are the weights of the policy before the update and $\epsilon$ is a hyperparamter to control the update of the policy. 

The reward function is a critical choice in RL scenarios. It encourages the RL agent to find a robust policy that, in our case, drives the physical system to the desired target state. In this work, we chose an instantaneous reward function of the fidelity $\mathcal{F}(t)$, computed between the current density matrix $\rho^{(c)}(t)$ and the target state, with 
\begin{equation}
r_t=-\log_{10}(1-\mathcal{F}(t)).
\end{equation}
Many stochastic trajectories (or "episodes"), starting from random coherent states, are run in parallel. Then, two networks $\pi_{\xi_1}$ and $V_{\xi_2}$ are optimized simultaneously. Please refer to ~\cite{schulman2017proximal} for detailed information. The neural network used is a multi-layer perceptron with two hidden layers of size 16 and 8, respectively. The reward decay is set as $\gamma=0.99$. The clip parameter is set as $\epsilon=0.2$. 

\section{Dicke state preparation in the absence of damping}
\label{App:gamma0}
\renewcommand\theequation{G.\arabic{equation}}

\begin{figure}[t]
\begin{center}
\includegraphics[angle=0,width=0.98\linewidth]{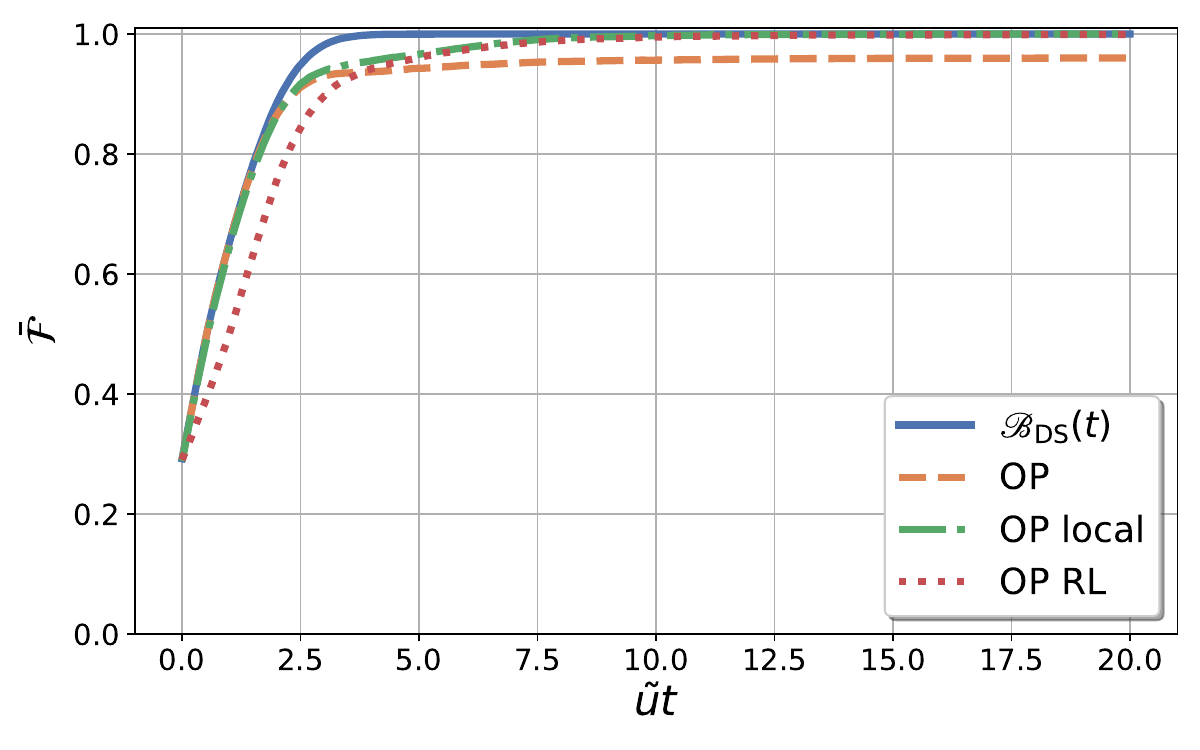} 
\end{center}
\caption{The fidelity, $\overline{\F}$(t), averaged over 10000 trajectories, to the maximally excited Dicke state as a function of time in the absence of collective damping ($
\gamma=0$). The system is prepared in a random spin-coherent state. $\mathscr{B}_{DS}(t)$ (blue, solid) represents the solution to the classical stochastic master equation in Eq.~\eqref{eq:CSME}. OP (orange, dashed) is calculated by applying the control strategy that locally minimizes the cost function in Eq.~\eqref{eq:Euc_cost}, OP local (green, dash-dotted) instead locally maximizes the fidelity. Finally, OP RL (red, dotted) is the strategy found by the DRL model.
Parameters: $N=4$, $\kappa/\tilde{u} = 0.4$, $m_T = N/2$.
}
\label{fig:gamma0}
\end{figure}

In Sec.~\ref{Sec:opt_cont} we demonstrated that the control strategies proposed in Eqs.~\eqref{eq:OP_OC} and \eqref{eq:FP_OC} are close to the optimal control strategies for maximizing $\overline{\mathcal{F}}_{ss}$ for OP and FP control respectively. This is further reinforced by the fact that in Fig.~\ref{fig:DickeN} (a) the optimal control strategy approaches $\mathscr{B}_{DS}$. Such convergence tells us that it is simultaneously possible to maximize the cost function in Eq.~\eqref{eq:Euc_cost} and keep the relative phases aligned as in Eq.~\eqref{eq:opt_align} (at least for $m_T = N/2$ and small $\kappa$). However, such alignment is lost in the $\gamma \to 0$ limit. This problem has been investigated before and multiple control strategies have been proposed~\cite{mirrahimi2007stabilizing,stockton2004deterministic}. 

In Fig.~\ref{fig:gamma0} we plot the average fidelity as a function of time for the control strategy in Eq.~\eqref{eq:OP_OC} compared with the DRL model and $\mathscr{B}_{\text{DS}}(t)$ which we define as the finite time solution to Eq.~\eqref{eq:CSME} which in the long time limit is equal to $\mathscr{B}_{DS}$. We initiate our system in a random coherent state and attempt to prepare the maximally excited state with $N=4$. We see that the OP strategy fails to converge to unit fidelity. This is because the relative phases lose their alignment and some trajectories become frustrated and fail to converge. The DRL on the other hand adapts by sacrificing some fidelity at early times to ensure that phase alignment remains sufficiently high to converge to the target state at later times. There are many ways to modify the control strategy to improve convergence, in Fig.~\ref{fig:gamma0} the line labeled ``OP local'' is obtained by choosing $\theta$ to locally minimize the cost function $\mathcal{C} = 1-\mathcal{F}$. $\mathscr{B}_{DS}(t)$ by construction always maintains perfect driving efficiency and therefore can always maximize driving speed towards the target state. We can see that the OP strategy follows $\mathscr{B}_{DS}(t)$ for early times but eventually the alignment diverges. $\mathscr{B}_{DS}(t)$ also beats the other strategies at all times, suggesting that it may also provide a bound on the control problem for finite times, not just the steady state as conjectured in the main text.

\bibliography{refs}
\end{document}